\documentclass[12pt,aps,amsmath,latexsym,amsfonts]{JHEP3}

\usepackage{epsfig}
\usepackage{amsfonts,amssymb,amsmath}
\usepackage[hang]{subfigure}

\newcommand{\be}{\begin{equation}}
\newcommand{\ee}{\end{equation}}
\newcommand{\bea}{\begin{eqnarray}}
\newcommand{\eea}{\end{eqnarray}}

\DeclareRobustCommand{\ceqref}{\eqref}

\title{Black holes as bubble nucleation sites}

\author{
Ruth Gregory$^{1,2}$\thanks{r.a.w.gregory@durham.ac.uk}, 
Ian G. Moss$^{3}$\thanks{ian.moss@newcastle.ac.uk}, 
Benjamin Withers$^{1,4}$\thanks{b.s.withers@soton.ac.uk} \\ 
$^1${\it Department of Mathematical Sciences and Centre for Particle 
Theory, South Road, Durham, DH1 3LE, U.K.}\\
$^2${\it Perimeter Institute, 31 Caroline Street North, 
Waterloo, ON, N2L 2Y5, Canada}\\
$^3${\it School of Mathematics and Statistics, Newcastle University, 
Newcastle Upon Tyne, NE1 7RU, U.K.}\\
$^4${\it Mathematical Sciences and STAG research centre, 
University of Southampton,
Higheld, Southampton SO17 1BJ, U.K.}
}

\abstract{
We consider the effect of inhomogeneities on the rate of false 
vacuum decay. Modelling the inhomogeneity by a black hole, we 
construct explicit Euclidean instantons which describe the 
nucleation of a bubble of true vacuum centred on the inhomogeneity.
We find that inhomogeneity significantly enhances the 
nucleation rate over that of 
the Coleman-de Luccia instanton -- the black hole acts as a 
nucleation site for the bubble. The effect is larger than 
previously believed due to the contributions
to the action from conical singularities.
For a sufficiently low initial mass, the original black hole is replaced 
by flat space during this process, as viewed by a single causal patch observer. 
Increasing the initial mass, we find a critical value above which a black hole
remnant survives the process. This resulting black hole can 
have a higher mass than the original black hole, 
but always has a lower entropy.  We compare
the process to bubble-to-bubble transitions, 
where there is a semi-classical Lorentzian description 
in the WKB approximation.
}

\keywords{vacuum decay, bubble nucleation, gravitational instantons}
\preprint{DCPT-13/43}

\begin{document}

\setlength{\baselineskip}{18pt}


\section{Introduction}


One of the most exciting aspects of quantum field 
theory is the possibility that the universe
can become trapped in a false vacuum state. For first 
order phase transitions, the decay rate
of the false vacuum state is exponentially suppressed,
\cite{coleman1977,callan1977}, allowing for a long-lived metastable state
with consequences for the very early universe, 
\cite{PhysRevD.23.347}, or the possible fate of the late universe,
\cite{1982Natur.298}.

Aside from cosmology, the decay rates for most commonly 
observed first order phase transitions
are greatly enhanced by the presence of nucleation sites 
for the preferred low temperature phase,
such as impurities or imperfections in the retaining walls.
The goal of this paper is to explore the cosmological version 
of a nucleation site by 
considering false vacuum decay in the presence of inhomogeneities. 
Enhancing the
transition rate could prevent the universe 
becoming trapped in a false vacuum state or
in a worst case scenario could bring about the premature end of the 
universe.

There have been very few previous investigations of 
the nucleation rates of true vacuum bubbles around black holes,
the closest to our work being Hiscock, \cite{PhysRevD.35.1161}. 
Berezin et al.\ investigated false vacuum decay around black 
holes in flat space, \cite{Berezin:1987ea}. 
The nucleation of a symmetric
phase around an evaporating black hole 
was investigated in \cite{PhysRevD.32.1333}. 
Some recent work has been done on false
vacuum decay due to modifications of the scalar field potential 
by the black hole, \cite{Cheung:2013sxa}.

A prototypical example of false vacuum decay, described 
by the Coleman-de Luccia (CDL) instanton, \cite{PhysRevD.21.3305}, 
takes place in the idealised setting of a maximally symmetric 
false vacuum universe -- de Sitter space-time. We will relax 
the initial condition of a homogeneous universe, and show 
that introducing inhomogeneity enhances the rate of production of 
true vacuum bubbles centred on the inhomogeneity. In particular, 
we consider the natural generalisation of the CDL Euclidean 
instanton solution to include the simplest form of inhomogeneity: 
a black hole. We are thus led to study the formation of 
vacuum bubbles in the false vacuum background described by the 
Schwarzschild-de Sitter black hole (SdS) metric:
\be
ds^2 = -f(r)dt^2 + \frac{dr^2}{f(r)} + r^2 d\Omega_2^2, 
\quad\qquad f(r) \equiv 1 - \frac{2GM}{r} - \frac{r^2}{\ell^2},
\label{sdscausal}
\ee
where the de Sitter radius, $\ell$, is related
to the energy density of the false vacuum, $\varepsilon$, by the
relation $\ell^2=3/(8\pi G{\varepsilon})$. As usual, the two positive roots 
of $f(r)$ correspond to the locations of the black hole horizon, $r_h$, 
and of the cosmological horizon, $r_c$. The horizons coincide when 
$GM/\ell= GM_N/\ell \equiv 1/\sqrt{27}$, which corresponds to
the Nariai solution, \cite{Nar}.

We shall construct the analogue of the 
thin-wall Euclidean `bounce' solution in the presence of 
finite mass, $M$. For convenience we work with the Euclidean 
section obtained by performing a Wick rotation, $t =- i \tau$, 
from the causal patch of SdS described by \eqref{sdscausal}, where 
we can take the coordinate $\tau$ to have period $\beta$. 
For an arbitrary choice of $\beta$ there are conical 
singularities at $r_h$ and $r_c$, the fixed points under the 
action of rotation in Euclidean time. In the thin-wall limit, 
an oscillatory bubble wall trajectory, $(\tau(\lambda), r(\lambda))$, 
describes the locus where we will match an exterior SdS solution with 
$r>r(\lambda)$ onto an interior true vacuum 
region with $r<r(\lambda)$. 
A sketch of the wall trajectory is shown in 
figure \ref{euclideancartoon}.
\FIGURE{
\includegraphics[width=0.7\textwidth]{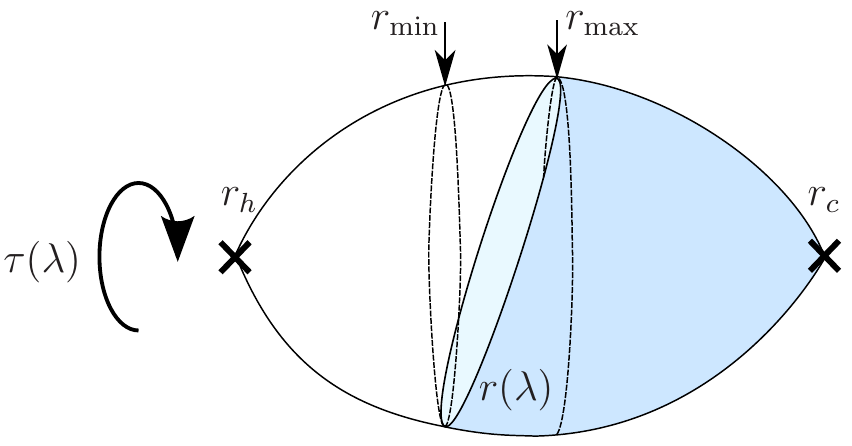}
\caption{Cartoon of a thin-wall in the Euclidean continuation 
of SdS. For clarity, we display only the $\tau,r$ surface,
and the $S^2$ is not shown. The shaded region indicates the false-vacuum 
exterior. The interior is not shown.  The time circle has been 
identified with the period of the wall solution shown, 
leading to conical singularities at the fixed points indicated 
by the black crosses.}
\label{euclideancartoon}
}
Note that because the Euclidean bubble wall trajectories are 
oscillatory in $\tau$ with their own characteristic period, 
$\beta_{wall}$, neither conical singularity can be removed with 
a choice for the period $\beta$ as we must choose $\beta = \beta_{wall}$. 
The field equations break down at the conical singularities, but 
despite this, a careful treatment of the conical singularities 
can be made, and the appropriate contribution to the Euclidean 
action computed. Finite action solutions with
conical singularities are known in the literature as 
singular instantons, \cite{Hawking:1998bn,Turok:1998he}. 

In the cases where $\partial_\tau$ is a Killing vector -- for example 
in the absence of the bubble wall, or for a $\tau$-independent bubble 
wall -- we show that the Euclidean action is always \emph{independent} 
of the choice of $\beta$ once the conical singularities are taken into 
account. This removes a possible ambiguity in the tunnelling rates 
when using singular instantons. Our formula for the
actions of symmetric singular instantons generalises the `NUTs and 
bolts' formula of Gibbons and Hawking, \cite{gh1979}.
For the non-static bubble, we find precise agreement for the value 
of the action with two cases where we are aware of an equivalent,  
manifestly regular construction: the CDL instanton with zero mass, and a 
second special case at finite $M_\ast$ where there is a critical, 
$\tau$-independent bubble wall, corresponding to solutions 
found in \cite{Garriga:2004nm} .

The main physical result of our analysis is the behaviour of the nucleation 
rate of true vacuum bubbles, $\Gamma(M)$, as a function of the 
initial black hole seed mass $M$. We find that there are two regimes, 
delineated by a critical mass,
\be
G M_C= \frac{32}{27} \frac{(8\pi G \sigma \ell)^3 }
{(4+(8\pi G \sigma \ell)^2)^2}, 
\label{MCintro}
\ee
where $\sigma$ is the tension of the thin-wall:
\begin{itemize}
\item For an initial seed black hole with $M<M_C$, the rate is an increasing 
function of the mass,
\be
\frac{\partial \Gamma(M)}{\partial M}\bigg|_\sigma >0,
\ee
with the dominant instanton corresponding to the nucleation of flat space 
inside the bubble. The nucleation rate for a 
bubble of true vacuum which replaces a finite mass SdS black hole is 
actually \emph{higher} than that of the CDL case. 
Although this conclusion agrees qualitatively with the previous work of 
Hiscock \cite{PhysRevD.35.1161}, there the contributions from 
conical singularities were not taken into account. The result here is a
far larger increase in the bubble nucleation rate. 
\item For an initial seed black hole with mass $M>M_C$, the rate is 
a decreasing function of the mass and for sufficiently large black holes 
the rate eventually becomes subdominant to that of the CDL instanton. 
For this range we find that the dominant process corresponds to bubble 
nucleation with a vacuum black hole, which we shall here term `the 
remnant'. The mass of this remnant black hole may 
be higher than $M$, but the horizon area is always smaller. 
\item The critical value, $M=M_C$, marks the point at which the 
remnant black hole becomes vanishingly small when approached 
from above, giving way to flat space inside
the bubble. Here, the transition rate is maximised, and is a 
decreasing function of the wall tension $\sigma$.
\end{itemize}

In contrast to the Euclidean instanton approach, in this paper we 
also consider the Lorentzian WKB method,
\cite{Farhi:1989yr,Fischler:1989se,Fischler:1990pk,Aguirre:2005xs,Aguirre:2005nt}, 
used to calculate the probability for transitions between bubble 
solutions: `bubble-to-bubble' transitions. These methods have 
been used mostly to investigate the nucleation of false vacuum 
bubbles in the context of creating baby universes, \cite{Farhi:1989yr}. 
The basic idea of the WKB method is to formulate an action 
which depends only on the bubble wall trajectory and then 
use the associated Schr\"odinger equation to calculate tunnelling 
probabilities. We present a new action for the bubble wall and 
show that in general, the bubble-to-bubble transition rate 
calculated via the WKB method is related to the spontaneous 
nucleation rate calculated in the singular instanton approach
via a numerical factor depending only on the black hole entropy.
We propose that this represents a type of crossing relation 
for the amplitude describing the bubble-to-bubble transition.

The paper is organised as follows. We begin in section 
\ref{euclideansec} with a computation of the thin-wall 
trajectories in both the Lorentzian, \eqref{sdscausal}, 
and Euclidean pictures. In section \ref{actions} we 
present a general derivation of the Euclidean action in the 
presence of conical singularities, which we then evaluate 
for the wall trajectories for generic values of the 
mass to compute nucleation rates in section \ref{sectionTunnelling},
where we also outline the dominant processes. In section \ref{wkb} 
we elucidate the connection to the Lorentzian WKB approach for 
bubble-to-bubble transitions. We conclude in section \ref{discussion}.

\section{Lorentzian and Euclidean thin-wall space-times}
\label{euclideansec}

We consider the following system of gravity and matter fields, 
\be
S = \frac{1}{16\pi G}\int_{\mathcal{M}}{\cal R}
+\int_{\mathcal{M}}{\cal L}_m(g,\phi),
\label{4Daction}
\ee
where the manifold $\mathcal{M}$ has a metric $g$ of 
Lorentzian signature $(-,+,+,+)$ and Ricci scalar ${\cal R}$. For 
each vacuum present in this theory we can construct a 
one parameter family of SdS black hole solutions \eqref{sdscausal}. 
The length scale $\ell$ is determined by the cosmological constant in the 
chosen vacuum, $\ell^2 = 3/\Lambda$. In this Lorentzian system 
\eqref{4Daction} we are interested in constructing space-times 
which describe two vacua separated by a thin bubble wall, and
as with CDL, we will use Israel's junction conditions, \cite{Israel},
to match a solution of the form \eqref{sdscausal} with mass $M_-$ 
and cosmological constant $\Lambda_-$ (the `inside') across a 
thin bubble wall of tension $\sigma$ to a solution mass $M_+$ 
and cosmological constant $\Lambda_+$ (the `outside'). 
These will in general be time-dependent bubble wall trajectories, 
many of which will correspond to a reflection or bounce.

We may also study Euclidean solutions obtained by performing 
the Wick rotation $t = -i \tau$ from the causal patch 
\eqref{sdscausal} to obtain a closed Euclidean manifold,
\be
ds^2 = f(r)d\tau^2 + \frac{dr^2}{f(r)} + r^2 d\Omega_2^2, 
\quad\qquad f(r) \equiv 1 - \frac{2GM}{r} - \frac{r^2}{\ell^2},
\label{sdseuclid}
\ee
which solve the equations of motion coming from the 
corresponding Euclidean action given by $I=-iS$. 
Proceeding by analogy with the CDL instanton, we 
may also construct a family of Euclidean thin-wall solutions 
separating different vacuum solutions of the form \eqref{sdseuclid}. 
We shall later make the interpretation that the on-shell 
action for these solutions, $I$, determines the rate of 
bubble nucleation, just as in the CDL case. Indeed 
when $M_-=M_+=\Lambda_-=0$ we obtain the CDL result, 
however, unlike Coleman and de Luccia, we will work entirely in 
the Euclidean continuation of a single causal static
patch of SdS i.e., \eqref{sdseuclid}. 

\subsection{Lorentzian bubbles}

In the thin wall description of the bubble, we describe the 
trajectory of the wall by local coordinates on each side of the wall:
\be
X^a_\pm = (t_\pm(\lambda), r_\pm(\lambda), \theta,\phi)
\ee
where for convenience, we take $\lambda$ to be the proper time
of an observer comoving with the wall,
\be
f_\pm(r_\pm)\dot{t}_\pm^2 - \frac{\dot{r}_\pm^2}{f_\pm(r_\pm)} =1\,.
\label{tequations}
\ee
The intrinsic coordinates on the wall are $\xi^A = (\lambda,\theta,\phi)$,
and the induced metric is
\be
ds^2 = -d\lambda^2 + r_\pm^2(\lambda) \left [ d\theta^2
+ \sin^2 \theta d \phi^2 \right]\,.
\label{inducedmetric}
\ee
Clearly, if the wall is to make physical sense as a boundary between
two regions, we require 
$r_+=r_- \equiv R(\lambda)$.

Next, we construct a normal one-form on each side of the wall
\be
n_\pm = \left(-\dot{r}_\pm dt_\pm + \dot{t}_\pm dr_\pm\right) 
\label{normal}
\ee
with the sign chosen so that it is always pointing towards increasing
$r$ for ${\dot t} >0$. From these, we construct the extrinsic
curvature of each side of the wall:
\be
K_{_\pm AB} = X^a_{_\pm,A} X^b_{_\pm,B}\nabla_a n_{_\pm b}\,.
\ee
Treating the wall's stress tensor $T^{w}_{ab}$ as a distributional 
source, we may construct the surface stress tensor,
\be
S_{ab} \equiv  \int T^{w}_{ab}dl,
\ee
and the Israel junction conditions, \cite{Israel}, then relate the 
energy-momentum of the wall to the geometry of its embedding 
measured via a jump in the extrinsic curvature across the wall:
\be
\Delta K_{ab}\equiv K_{_+ab}-K_{_-ab} 
= - 8\pi G\left(S_{ab} - \frac{1}{2}h_{ab}S\right)
\label{junctioncondition}
\ee 
For a wall of tension $\sigma$ we 
have $S_{ab}=-\sigma h_{ab}$ and this equation reduces to 
\be
\frac{1}{R}\left(f_+(R)\dot{t}_+-f_-(R)\dot{t}_-\right) = - 4\pi G \sigma.
\label{Israeleq}
\ee
Using \eqref{tequations}, we can substitute for ${\dot t}_\pm$ on
each side, and rearranging reveals a Friedman-like equation,
\cite{BCG}, for the trajectory of the wall
\be
\left(\frac{\dot{R}}{R}\right)^2 = \bar{\sigma}^2 -\frac{\bar{f}}{R^2}
+ \frac{(\Delta f)^2}{16R^4\bar{\sigma}^2}.
\label{radialeom}
\ee
where $\bar{\sigma} \equiv 2\pi G \sigma$, $\bar{f} \equiv (f_-+f_+)/2$, 
and $\Delta f \equiv f_+-f_-$.  To completely determine the wall trajectory,
we must determine the time coordinate evolution, obtained
by combining \eqref{radialeom} with \eqref{tequations},
ensuring consistency with the Israel junction equation
\eqref{Israeleq}:
\be
f_\pm {\dot t}_\pm = \sqrt{f_\pm + {\dot R}^2}
= \mp {\bar\sigma}R - \frac{\Delta f}{4{\bar\sigma}R} 
\label{tdotexp}
\ee
\eqref{radialeom} and \eqref{tdotexp} now
completely describe the bubble wall trajectory 
for general $M_\pm,\Lambda_\pm$. 

In this paper we are primarily interested in the effects of mass on 
false vacuum decay, and for this purpose we 
consider the class of solutions where the interior solution is 
true vacuum, i.e.\ $\Lambda_-=0$, and the 
exterior is false vacuum, $\Lambda_+ = 3/\ell^2$. This 
includes the CDL case, which simply has $M_+ = M_-= 0$. The 
radial equation \eqref{radialeom} in this case becomes,
\be
\left(\frac{\dot{R}}{R}\right)^2 + \frac{1}{R^2} = \left(\bar{\sigma} 
+ \frac{1}{4\bar{\sigma}\ell^2} + \frac{G\Delta M}{2\bar{\sigma}R^3}
\right)^2 + \frac{2GM_-}{R}
\label{mplusminuseqn}
\ee
from which we may identify an effective potential governing the wall position, 
\be
2 U(R) =1 - \left(\frac{R}{\gamma} 
+\kappa_2 \frac{\gamma^2}{R^2}\right)^2 
-\kappa_1\frac{\gamma}{R}
\label{Upotential}
\ee
where we have introduced the parameters
\be
\gamma = \frac{4\bar{\sigma}\ell^2}{4\bar{\sigma}^2\ell^2+1}\,,
\qquad\kappa_1 = \frac{2GM_-}{\gamma}\,,\qquad
\kappa_2 = \frac{G\Delta M}{2\bar{\sigma}\gamma^2}\,.
\ee
It is clear that the overall qualitative nature of the solution depends 
only $\kappa_1$ and $\kappa_2$ with solutions at different values 
of $\gamma$ reached under simultaneous $\lambda$ and $R$ rescalings. 
Note however that the value of $\bar\sigma$ is important in 
determining ${\dot t}_\pm$.
The potential is qualitatively similar for varying $\kappa_1, \kappa_2$,
and is illustrated in figure \ref{potentialplot} for $\kappa_1=0$. 
For fixed $\kappa_2$, switching on $\kappa_1$ has a similar effect to
increasing $\kappa_2$, in that the potential is lowered, and the range of
disallowed ${\tilde R} = R/\gamma$ is decreased.
\FIGURE{
\includegraphics[width=0.6\textwidth]{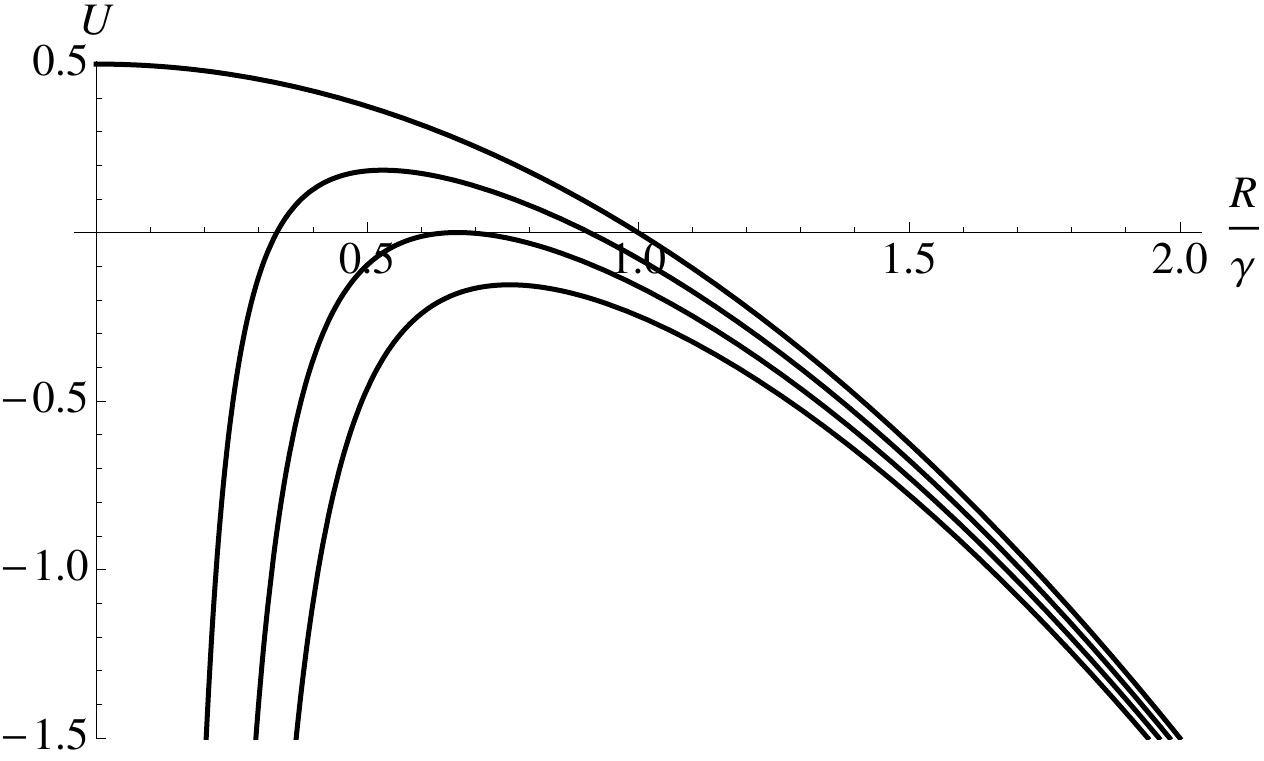}
\caption{
The potential $U$ shown here with $\kappa_1=0$. From top to bottom 
we show $\kappa_2 = 0, \frac{1}{2}\kappa_\ast,\kappa_\ast,\frac{3}{2}
\kappa_\ast$, where $\kappa_\ast=4/27$ marks the transition from 
bounce to transmission, with an unstable fixed-$r$ solution when 
$\kappa_2=\kappa_\ast$.}
\label{potentialplot}
}

For any value of $\kappa_2$, there is a critical value of $\kappa_1$
for which the maximum of the potential is at zero, delineating bouncing 
solutions from transmission solutions. This critical value can be obtained
by simultaneously solving $U=U'=0$, and leads to a maximal value 
$\kappa_1^\ast(\kappa_2)$, derived in
appendix \ref{kappalimits} (Eq.\ \eqref{kappa1star}).
Physically, we are restricted to $\kappa_1\geq0$, however, we see
from appendix \ref{kappalimits} that we can have $\kappa_2<0$, or a 
``bigger on the inside'' mass with a regular bounce solution.
The only constraint on $\kappa_{1,2}$ is that they allow for a range
of $\tilde R$ for which $U$ is positive, determined by 
\eqref{kappaconstraints}, and also that ${\dot t}_\pm \geq0$,
which ensures we have a positive tension properly
oriented wall, and determines a minimum value for $\kappa_1$
(Eq.\ \eqref{kappamin}). 
 
Because we will ultimately be interested in Euclidean wall trajectories, 
we shall consider only those bubble solutions which reflect. 
For certain parameters we can construct the solutions 
analytically. For example, at $\kappa_1=\kappa_2 =0$ we have,
\bea
R(\lambda) &=& |\gamma| \cosh\left(\frac{\lambda}{\gamma}\right) 
\label{lcdl1}\\
t_-(\lambda) &=& \gamma \sinh \left(\frac{\lambda}{\gamma} \right)
\label{lcdl2}\\
\sqrt{\ell^2-\gamma^2} \tanh\left ( \frac{t_+(\lambda)}{\ell} \right )
&=& \gamma \sinh\left( \frac{\lambda}{\gamma}\right)
\label{lcdl3}
\eea
which parametrically describes the trajectory, $\cosh^2\left(
\frac{t_+}{\ell}\right) \left(r^2-\ell^2 \tanh^2\left ( 
\frac{t_+}{\ell}\right)\right) =\gamma^2$. To see that this 
solution is just the CDL solution, \cite{PhysRevD.21.3305}, but in 
the causal patch, we make the coordinate transformation 
$r(\rho,\chi), t_+(\rho,\chi)$, with
\be
\cosh^2\left(\frac{t_+}{\ell}\right) 
\left(r^2-\ell^2 \tanh^2\left( \frac{t_+}{\ell}\right) \right) 
= \ell^2 \sin^2\left( \frac{\rho}{\ell}\right) , \;\;\; r^2 
= \ell^2 \sin^2\left( \frac{\rho}{\ell} \right) \sin^2\chi
\ee
which results in the metric on a round S$^4$ 
\be
ds^2 = d\rho^2 + \ell^2 \cos^2\left (\frac{\rho}{\ell}\right) 
\left(d\chi^2 + \sin^2\chi d\Omega_2^2\right)
\ee
with the bubble wall sitting at fixed $\rho$, specifically, 
$\ell^2 \cosh^2\left (\frac{\rho}{\ell}\right) = \gamma^2$,
which agrees with \cite{PhysRevD.21.3305}.\footnote{To 
compare with \cite{PhysRevD.21.3305} we note some 
notational differences, we have $\bar{\rho}_{CDL} 
= \gamma$, $S_{1,CDL} = \sigma$, $\kappa_{CDL} = 8\pi G$ and 
$\varepsilon_{CDL} =3/(8\pi G\ell^2)$.}
Note however that \eqref{tdotexp} requires $2{\bar\sigma}\ell<1$
for ${\dot t}_+\geq0$, in essence stating that the CDL bubble wall
must remain within the static patch. Clearly one does not have this
restriction in the original CDL approach, however, for $2{\bar\sigma}\ell
>1$, the bubble has now consumed more than half of the de Sitter
hyperboloid. 

A second family of analytic solutions can be found when 
$\kappa_1$ and $\kappa_2$ satisfy \eqref{kappaconstraints},
$\kappa_1=\kappa_1^\ast(\kappa_2)$,
for which there is a critical, unstable bubble wall solution sitting at the 
maximum of the potential, $R_\ast=2^{-2/3} \gamma \left ( \kappa_1^\ast
+ 2 \kappa_2 + \sqrt{\kappa_1^{\ast2} + 4 \kappa_1^\ast
\kappa_2 +36\kappa_2^2} \right )^{1/3}$. Finally, for the 
remaining solutions\footnote{The range of $\kappa_1$ can be 
modified if $\kappa_2<0$, as we discuss in the next subsection, 
\textsection\ref{euclideanbubbles}.}, $0<\kappa_1< \kappa_1^\ast$, 
we find that there are 
two solutions, corresponding to small-$r$ and large-$r$ bounces.  
The former corresponds to a bubble entering through the past black 
hole horizon, growing, turning around and then falling through the 
future black hole horizon, whilst the latter corresponds to the same 
process but for the cosmological horizon; this is easily seen from the 
potential. We have explicitly constructed examples of these trajectories 
numerically.

\subsection{Euclidean bubbles\label{euclideanbubbles}}

The effect of performing a Wick rotation $t=-i\tau$ can be 
described in the equations of the last section by simultaneously 
Wick rotating the world-volume proper time, $\lambda = -i x$. The 
overall effect of this on the equation governing the bubble wall 
position is a flip in the sign of the potential, \eqref{Upotential}. 
Thus we obtain solutions only in the parameter range 
for which Lorentzian bounce solutions exist.
\FIGURE{
\includegraphics[width=0.89\textwidth]{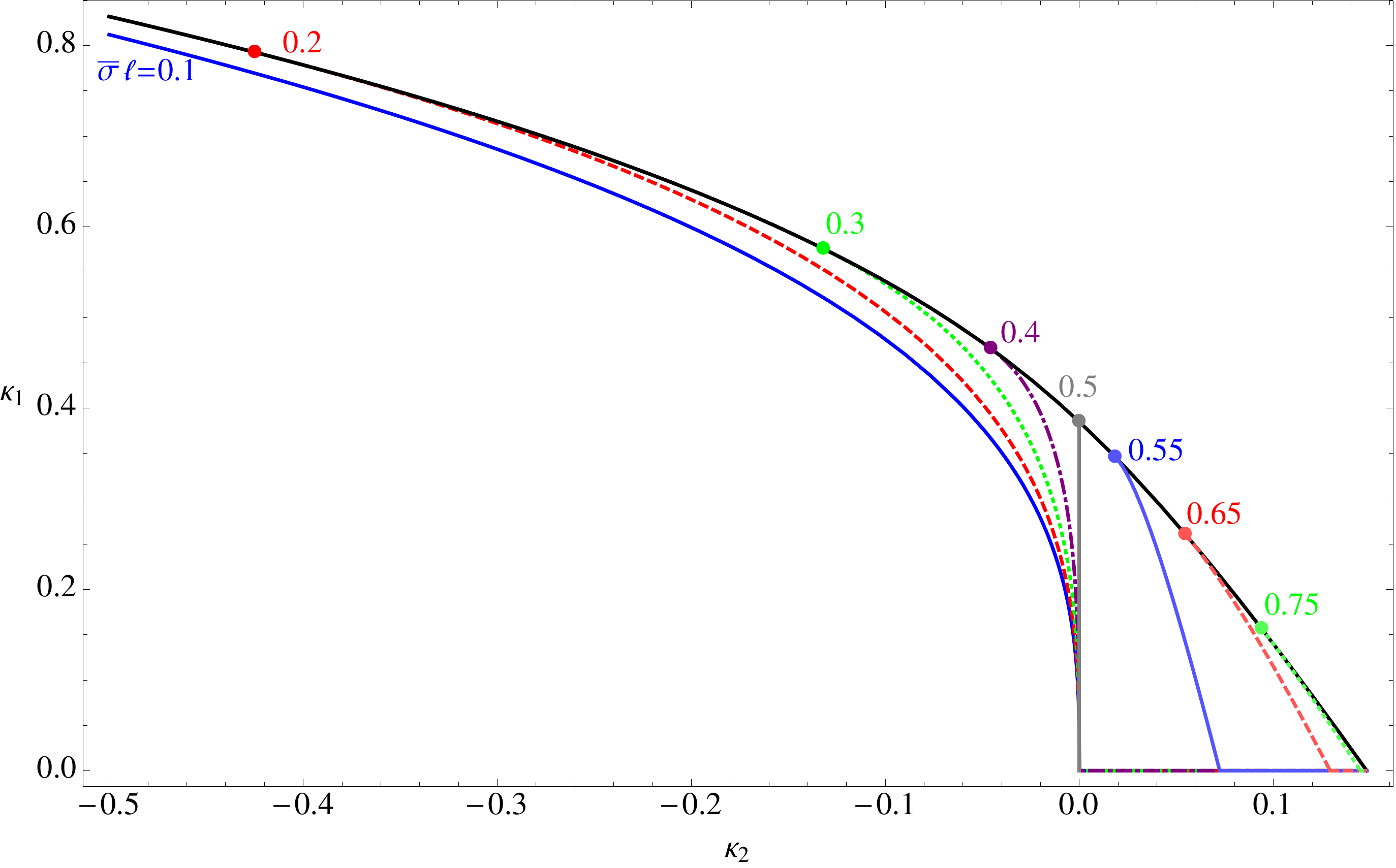}
\caption{
The allowed range of $\kappa_2$ and $\kappa_2$ for ${\bar\sigma}\ell\leq
\sqrt{3}/2$. 
The upper solid black line shows the maximum value of $\kappa_1$, 
$\kappa_1^\ast$, for a given $\kappa_2$.
The lower lines show the minimum value allowed for $\kappa_1$ at
various values of ${\bar\sigma}\ell$. 
From left to right: ${\bar\sigma}\ell=0.1$ in solid blue, $0.2$ in dashed red, 
$0.3$ in dotted green, $0.4$ in dot-dash purple, $0.5$ in grey, 
$0.55$ in light blue, $0.65$ in light dashed red, and $0.75$ in light
dotted green. The data points in matching
color indicate the limiting value of $\kappa_2$ for each ${\bar\sigma}\ell$.
The allowed range closes off entirely above $\bar\sigma\ell=\sqrt{3}/2$.
}
\label{fig:kappas}
}

In appendix \ref{kappalimits}, we detail the constraints on $\kappa_1,
\kappa_2$ for a regular instanton to exist. Briefly, for a bounce, we 
require $U\geq0$ for some range of $\tilde R$, leading to an upper 
bound $\kappa_1 \leq \kappa_1^\ast$, given in \eqref{kappaconstraints}. 
However, a regular instanton also requires positivity of
${\dot \tau}_+$, which leads to a lower bound $\kappa_1
\geq \kappa_{1,min}$, as derived in \eqref{kappamin}.
Figure \ref{fig:kappas} shows the allowed parameter ranges of $\kappa_1$
and $\kappa_2$ for a selection of values of ${\bar\sigma}\ell$. 
The range of $\kappa_1$ is shown as a function of $\kappa_2$,
the maximum $\kappa_1^\ast$ being shown as a solid black line. 
For a selection of ${\bar\sigma}\ell$, the minimum value of $\kappa_1$ is
shown, as well as the limiting value where the range of $\kappa_1$ 
eventually closes off, i.e.
\be
\kappa_2 = \frac{-(1-2{\bar\sigma}\gamma)\ell^3}{3 \sqrt{3} \gamma^3} \; , 
\;\; \kappa_1 = \frac{(3-4{\bar\sigma}^2\ell^2)\ell}{3 \sqrt{3} \gamma} \;\; 
\Rightarrow \;\; GM_+ = GM_N \;,\;\; R_\ast = \frac{\ell}{\sqrt{3}}
\label{kappalimit}
\ee
For $\kappa_2 \le -(1-2{\bar\sigma}\gamma)\ell^3/(3 \sqrt{3} \gamma^3)$,
there is no well defined solution, and it follows that there is an upper limit 
to the tension of $\bar\sigma\ell<\sqrt{3}/2$. On the other hand, for any
large negative $\kappa_2$ we can always find a $\bar\sigma$ small
enough to allow for a bounce. 

When $\kappa_{1,min} <\kappa_1 <\kappa_1^ \ast$ there are a pair of
Lorentzian bounce solutions, and correspondingly there is 
a single Euclidean bubble wall which is periodic in $x$ with the same 
turning points. Note that the periodicity in the time coordinates on each
side of the Euclidean wall need not be, and indeed in general is not, the
same. Again, we have constructed these solutions numerically.

When $\kappa_1=\kappa_2=0$ corresponding to the CDL case, 
there is only a single Lorentzian bounce solution. In the Euclidean picture the 
bubble begins at $R=0$ when $x = -\gamma\pi/2$,  grows 
to its maximum size at the turning point, and then retreats 
once more to $R=0$ when $x = \gamma\pi/2$. This can 
be seen by Wick rotating the Lorentzian solution \eqref{lcdl1}--\eqref{lcdl3}. 
Notice that this interval in the external Euclidean time coordinate, 
$\tau_+$, runs over a particular range, $\beta_{\text{CDL}}$, which 
does not correspond to the regularity condition at the cosmological 
horizon, in general. We also note that $\beta_{\text{CDL}} 
= \lim_{\kappa_1,\kappa_2\to 0}\Delta\tau_+$, so it is continuously 
connected to the massive oscillatory solutions in 
an appropriate sense.

\section{Computing the Euclidean action}
\label{actions}

In the last section we computed Lorentzian and Euclidean thin-wall 
bubble trajectories. The Euclidean trajectories oscillate between the 
turning points of the corresponding Lorentzian solutions. In this 
section we compute the Euclidean on-shell action for the bubble-wall 
solution $I$, and for no bubble-wall $I_{SDS}$, from which we calculate
\be
\Gamma \propto e^{-B}, \quad \text{where}\quad B 
= I - I_{\text{SdS}}.
\label{decayrate}
\ee
We claim that $\Gamma$ gives the rate at which bubbles are 
nucleated in a false vacuum SdS universe, centred on the black hole.
Similar claims are made for calculation of the decay rate of the 
false vacuum in de Sitter space, \cite{PhysRevD.21.3305},  
for black hole nucleation, \cite{Mellor:1989gi,Mellor:1989wc,Dowker:1993bt}, 
and for open universe nucleation, \cite{Hawking:1998bn,Turok:1998he}. 
The only case where the formula has a rigorous justification
is in flat space, \cite{coleman1977}. An interesting interpretation 
of the CDL instanton with some support
for the formula has been given in \cite{Brown:2007sd}.
Nevertheless, the use of instantons to calculate nucleation
rates in curved space has to be treated as speculative, and the 
results considered with a degree of caution.

\subsection{General results}

It is instructive to first consider the case where the Euclidean 
space $\cal M$ has a Killing vector, $\partial_\tau$. This can 
occur in the absence of a bubble wall, or for a $\tau$-independent 
wall configuration. As previously noted, $\cal M$ will in general 
contain a finite number of conical singularities, we also allow 
for a general scalar field in the space-time, provided it satisfies
the required background symmetries. 

The contributions from the conical deficits are determined by 
isolating them within a small region around each,
$\mathcal{B}_i= \{x^\mu : |r-r_i| < {\cal O} (\epsilon^2) \}$,
smoothing out the conical deficit, performing our integral,
then sending $\epsilon\to0$.
Although in general, one cannot regulate a co-dimension two 
$\delta-$function singularity in general relativity, \cite{GT}, 
for the particular case of a product metric, the limit is well-defined, 
as the ambiguity occurs due to nontrivial physical content in the 
transverse components of the energy-momentum tensor, which 
are not present in the special case of the product metric.
We therefore write the Einstein-Hilbert action as
$I=I_{\mathcal{M}-\mathcal {B}}+I_ \mathcal {B}$, where
\begin{eqnarray}
I_{\mathcal{M}-\mathcal {B}}&=&-\frac{1}{16\pi G}
\int_{\mathcal{M}-\mathcal {B}}{\cal R}-
\int_{\mathcal{M}-\mathcal {B}}{\cal L}_m(g,\phi)
+\frac{1}{8\pi G}\int_{\partial \cal B} K\\
I_{\cal B}&=&-\frac{1}{16\pi G}
\int_{\cal B} {\cal R}+\frac{1}{8\pi G}\int_{\partial \cal B} K
\end{eqnarray}
where the appropriate Gibbons Hawking boundary terms have
been added (with {\it inward} pointing normals)
at each ball boundary.

To evaluate this on-shell we perform a foliation of ${{\cal M}- {\cal B}}$ 
with a family of surfaces $\Sigma_\tau$ (assuming the global topology 
permits), with $0<\tau<\beta$. 
For this foliation we introduce coordinates with lapse $N$ and 
shift functions $N^i$, as well as the induced metric ${}^{(3)} g_{ij}$, 
its conjugate momentum $\pi^{ij}$, and conjugate momentum of 
the matter field, $\pi$. The leaves of the foliation have 
boundaries at the ends, and the canonical decomposition of 
such foliations has been 
investigated by Hawking and Horowitz, \cite{Hawking:1995fd}. 
The first piece of our action becomes, 
\be
I_{{\cal M} - {\cal B}}=\frac{1}{16\pi G}\int_0^\beta\,d\tau\left[
\int_{\Sigma_\tau} \left({}^{(3)}\partial_\tau g_{ij}\pi^{ij}
+\partial_\tau\phi\,\pi-N{\cal H}-N^i{\cal H}_i\right)
-\int_{\partial {\cal B}_\tau}N k\right],
\ee
where $\cal H$ and ${\cal H}_i$ are the Hamiltonian and momentum 
constraints and $k$ is the extrinsic curvature of 
$\partial {\cal B}_\tau\equiv\partial {\cal B}\cap\Sigma_\tau$. 
We have ${\cal H}={\cal H}_i=0$ and furthermore the symmetry 
implies $\partial_\tau\phi={}^{(3)}\partial_\tau g_{ij}=0$. The 
contribution from the Gibbons-Hawking term is sub-leading in the 
expansion about the conical singularity, and hence $Nk=O(\epsilon)$, 
i.e.\ $I_{{\cal M} - {\cal B}}=0$ to leading order. This possibly
surprising result is readily confirmed by a direct computation of the 
action $I_{{\cal M} - {\cal B}}$ in the case of a pure cosmological 
constant `dark energy' source.

Turning to the contribution from the conical singularities, we show
in the appendix that the contribution from a single conical defect
region $B_i$ is given by its area ${\cal A}_i$,
\be
-\int_{B_i} {\cal R}+2\int_{\partial B_i} K=-4\pi {\cal A}_i\,.
\ee
Putting all the contributions to the action together gives
\be
I=-\frac{1}{4G}\sum_i{\cal A}_i.\label{killingI}
\ee
At first sight, this appears to be an example of the classic  `NUT's and 
bolts'  formula for the action of a gravitational
instanton due to Gibbons and Hawking, \cite{gh1979}, but the difference 
is that we have extended the result to singular instantons.  
The remarkable feature is that the conical deficit angle
does not appear in the action, which is explicitly independent
of the period $\beta$. As a 
consequence, a possible source of ambiguity in using singular instantons
has been annulled. We shall  employ this result in order to compute 
the Euclidean on-shell action for SdS and static bubbles, although 
the result applies in any dimension, and can be further extended 
to rotating or charged spaces without any difficulty.

Solutions with a moving bubble wall break the time-translation
symmetry of the full space-time, but the result can be extended 
if the geometries on both sides of the bubble wall still individually 
possess the Killing vector, $\partial_\tau$. 
Consider the spherically symmetric metric
\be
ds^2=f(r)d\tau^2+f(r)^{-1}dr^2+r^2d\Omega^2,
\label{thesamemetric}
\ee
where now we allow two conical defects at $r_h$ and $r_c$, and (in principle) 
a more general form for $f(r)$ than used previously. There is a 
wall whose location is parametrised by $r=R(\lambda)$ 
and $\tau(\lambda)$, as illustrated in figure \ref{euclideancartoon}. 
Let ${\cal M}_\pm$ denote the regular parts of the manifold to the 
right and left of the wall, regions $\cal B$ covering the conical 
defects as before, and $\cal{W}$ the contribution of the wall itself 
and split the action into contributions from each region,
\be
I=I_{\cal B}+I_-+I_++I_{\cal W},
\ee
where $\cal B$ covers the conical defects as before,
\be
I_{\cal W}=-\int_{{\cal W}}{\cal L}_m(g,\phi)
= \int_{{\cal W}}\sigma
\ee
is the action of the thin wall, and
\be
I_\pm=-\frac{1}{16\pi G}\int_{{\cal M}_\pm}{\cal R}
-\int_{{\cal M}_\pm}{\cal L}_m(g,\phi)
+\frac{1}{8\pi G}\int_{\partial {\cal M}_\pm} K.
\ee
are the remaining bulk actions with the relevant Gibbons-Hawking
boundary terms. Note $\partial{\cal M}_\pm$ include both the boundaries
at the conical deficit excision balls, as well as the boundary on each 
side of the wall. As is conventional, these boundary integrals are
evaluated with {\it inward} pointing normals, which means that on
the inner wall boundary $r_- =R$, this normal will in fact have the 
{\it opposite} sign to the one usually used in the computation of the
Israel junction conditions, and therefore there will be an apparent 
sign difference when we come to use that substitution, which is simply due
to this vexatious disparity in conventions.

In order to decompose the action into space and Euclidean time 
we use the identity
\be
{\cal R}={}^3{\cal R}-K^2+K_{ab}^2-2\nabla_a(u^a\nabla_b u^b)
+2\nabla_b(u^a\nabla_a u^b), \label{curvatureidentity}
\ee
where the vector $u^\mu$ is normal to $\Sigma_\tau$ \cite{Hawking:1995fd}. 
After integration by parts, and taking the conical deficit excision radius
$\epsilon\to0$, we obtain
\be
\begin{aligned}
I_\pm=&-\frac{1}{16\pi G}\int_0^\beta\,d\tau
\int_{\Sigma_\tau}
\left({}^3{\cal R}-K^2+K_{ab}^2-16\pi G{\cal L}_m\right)\\
&\qquad -\frac{1}{8\pi G}\int_{\cal W}K_\pm
+\frac{1}{8\pi G}\int_{\cal W} n_{_\pm b} u^a\nabla_a u^b,
\end{aligned}
\ee
with $n_{_\pm a}= \pm( \dot{\tau} dr-\dot{r}d\tau)$ the inward 
pointing one-form normal to $\cal W$ as described above.
The first integral reproduces the canonical action we had previously, 
and vanishes due to the killing symmetry and the constraints. 
The second term represents the integration of the singular part
of the Ricci scalar due to the thin wall, and by Israel's junction
condition (remembering the sign disparity) the extrinsic
curvatures on each side of the wall are related via $K_+ =
- 4 \pi G S - K_- =12\pi G\sigma - K_-$ for the surface
stress tensor $S_{ab} = -\sigma h_{ab}$. The final contribution is a 
boundary term coming from the wall, and $u^a\nabla_a 
u^b\, \partial_b=- f' \partial_r/2$ giving,
\be
I_\pm=\mp\frac{1}{16\pi G}\int_{\cal W}  f_\pm'  \dot{\tau}_\pm,
\ee
where the integrand is evaluated using the metric components 
on the appropriate side of the wall. 

Pulling all these pieces together with the previous result for 
the conical defects we reach our final result,
\be
I=-\frac{1}{4G}\left({\cal A}_h+{\cal A}_c\right)-\frac12\int_{\cal W}\sigma
-\frac{1}{16\pi G}\int_{\cal W}
\left(f_+'\dot{\tau}_+- f_-' \dot{\tau}_-\right),
\label{generalaction}
\ee
for the action of a space-time with a bubble wall separating two
regions of possibly different black hole masses and effective
cosmological constants.

\section{Tunnelling from the false vacuum}
\label{sectionTunnelling}

Having demonstrated how to calculate the action
of a singular instanton, we would like to apply the method to the
situation of tunnelling catalysed by a ``point source impurity'' -- the
black hole. In general we can consider the case where a remnant black hole
remains in the true vacuum after the nucleation process, possibly with a different
mass from the original black hole. We shall consider the general case
in section \ref{insideblackhole}. First however, we
consider some special cases,  where the seed black hole is wiped out during the tunnelling 
process, leaving no remnant black hole within the interior of the bubble, or where
the remnant black hole corresponds to the `static' bounce.

For future reference, in the case of tunnelling from SdS to general 
Schwarzschild ($\Lambda_-=0$, $\Lambda_+=\Lambda$, general $M_+, M_-$),
we note the expressions for the (Euclidean) wall
trajectory 
\be
{\dot R}^2 = 1 - \frac{2GM_-}{R} - \left ( \frac{R}{\gamma} 
+ \frac{G\Delta M}{2 {\bar \sigma} R^2} \right )^2
\label{eucReom}
\ee
and the action 
\be
I_{\kappa_1,\kappa_2}=\frac{1}{4G}\left [ -\left({\cal A}_h+{\cal A}_c\right)
+\int d\lambda \left[\left ( 2R - 6GM_+\right)\dot{\tau}_+
- \left (2R-6GM_-\right) \dot{\tau}_-\right]\right]
\label{mplusminusaction}
\ee
where we have used the euclidean Israel equation to substitute for
$\bar\sigma$, and for clarity of presentation we have labelled the action 
using subscripts for the parameters $\kappa_1$ and $\kappa_2$.
We consider the two analytic cases first, 
then discuss other cases numerically.

\subsection{Coleman de Luccia bubbles}
\label{cdlsec}

The first case we consider in the nucleation of a Minkowski region
in de Sitter space through the CDL process, where the masses
$M_+=M_-=0$. Recall that the Lorentzian CDL bubble in the static patch is 
parameterised by \eqref{lcdl1}--\eqref{lcdl3}, which gives the
Euclidean bubble:
\be
R= \gamma \cos\left(\frac{\lambda}{\gamma}\right) \;,\;\;
t_- = \gamma \sin \left(\frac{\lambda}{\gamma}\right)\;,\;\;
t_+ = \ell\, {\rm arctan}\left [ 
\frac{\gamma \sin\left( \frac{\lambda}{\gamma}\right)}
{\sqrt{\ell^2-\gamma^2} }\right]
\label{ECDL}
\ee
Now, we take the integration in \eqref{mplusminusaction} to cover 
a single interval for which $R>0$, i.e.\ the 
wall traverses the interval $ -\gamma \pi/2\le\lambda\le\gamma \pi/2$. 
Notice that this interval in the external Euclidean time coordinate, 
$\tau_+$,  runs over a particular range, $\beta_+$, which does not 
correspond to the regularity condition at the cosmological horizon 
in general. 

Inputting the functions \eqref{ECDL} into \eqref{mplusminusaction} gives
\bea
\int d\lambda R\left (\dot{\tau}_+ -\dot{\tau}_-\right)
= \pi \left [ 2 \ell^2 - \gamma^2 - 2\ell \sqrt{\ell^2-\gamma^2}\right],
\eea
requiring tensions $4\bar{\sigma}^2\ell^2<1$. We also have
${\cal A}_h=0$ and ${\cal A}_c=4\pi \ell^2$, leaving us with
\be
I_{0,0} =  -\frac{\pi \ell^2}{G}\frac{1+8\bar{\sigma}^2\ell^2}
{(1+4\bar{\sigma}^2\ell^2)^2}.
\ee
The exponent $B= B_\text{CDL} $ and the associated tunnelling 
rate can then be obtained simply from \eqref{decayrate}, 
with $I_{\text{SdS}} = I_{\text{dS}}$ in this case, 
\be
B_\text{CDL} = \frac{\pi\ell^2}{G} \frac{16\bar{\sigma}^4\ell^4}
{\left(1+4\bar{\sigma}^2\ell^2\right)^2}
\ee
in agreement with the non-singular instanton 
calculation, \cite{PhysRevD.21.3305}.

\subsection{The critical static bubble wall with $M_-=0$}
\label{criticalresults}

The second analytic solution we considered in
Section \ref{euclideansec} was the critical case with 
$\kappa_1=\kappa_1^*(\kappa_2)$. When $M_-=0$,
this corresponds to the special choice $\kappa_2=\kappa_\ast=4/27$ 
which  is when there exists a single unstable Lorentzian bubble 
wall at fixed radius $R=2\gamma/3$. We may 
calculate the action using the integrals in \eqref{mplusminusaction}, where 
the result is independent of the integration range, or we 
may simply use the result \eqref{killingI} to obtain in both cases,
\be
I_{0,\kappa_\ast} =  -\frac{\pi r_c^2}{G}.
\ee
The exponent $B$ and the tunnelling rate can be obtained 
simply from \eqref{decayrate}, specialising $I_{\text{SdS}}$ 
to $\kappa_2=\kappa_\ast$ space-times. We find 
that in this case,
\be
B_\ast=\frac{\pi r_h^2}{G}\label{bstar}
\ee
Significantly, we find that $B_\ast<B_\text{CDL}$ when compared 
at the same tension, $\bar{\sigma}$.

\subsection{General $M_-=0$ bubbles \label{generalm0bubbles}}

Away from the special cases discussed above we must solve 
the wall trajectory equations \eqref{eucReom} and \eqref{tdotexp} 
numerically. As with the CDL case, when evaluating the integrals in 
the action \eqref{mplusminusaction}, we integrate over one period of the Euclidean 
wall trajectory. The instanton actions computed, $I$, are used 
to evaluate the exponent $B$ in the decay rate \eqref{decayrate}. 
We then compare $B$ with its $M=0$ value, $B_\text{CDL}$, at 
fixed tension. A sample of results are shown in figure \ref{bfig}. 
In all cases with finite $M$, as with the critical bubble wall 
case in section \ref{criticalresults}, we find
\be
B<B_\text{CDL}
\ee
when compared at fixed tension, $\bar{\sigma}\ell$. In particular, 
we observe that $B$ is a monotonically decreasing function of 
$M$ at fixed $\bar{\sigma}\ell$, hence the nucleation 
rate is monotonically increasing with the mass of the black hole, $M$.

Figure \ref{bfig} also shows that that there is no lower limit to the 
ratio $B/B_{CDL}$. This is in contrast to the results of
Hiscock \cite{PhysRevD.35.1161}, who gave an lower bound of around 0.6. 
The difference is due to the fact that Hiscock did not include the 
contribution to the action from the conical singularities.
\FIGURE{
\includegraphics[width=0.7\textwidth]{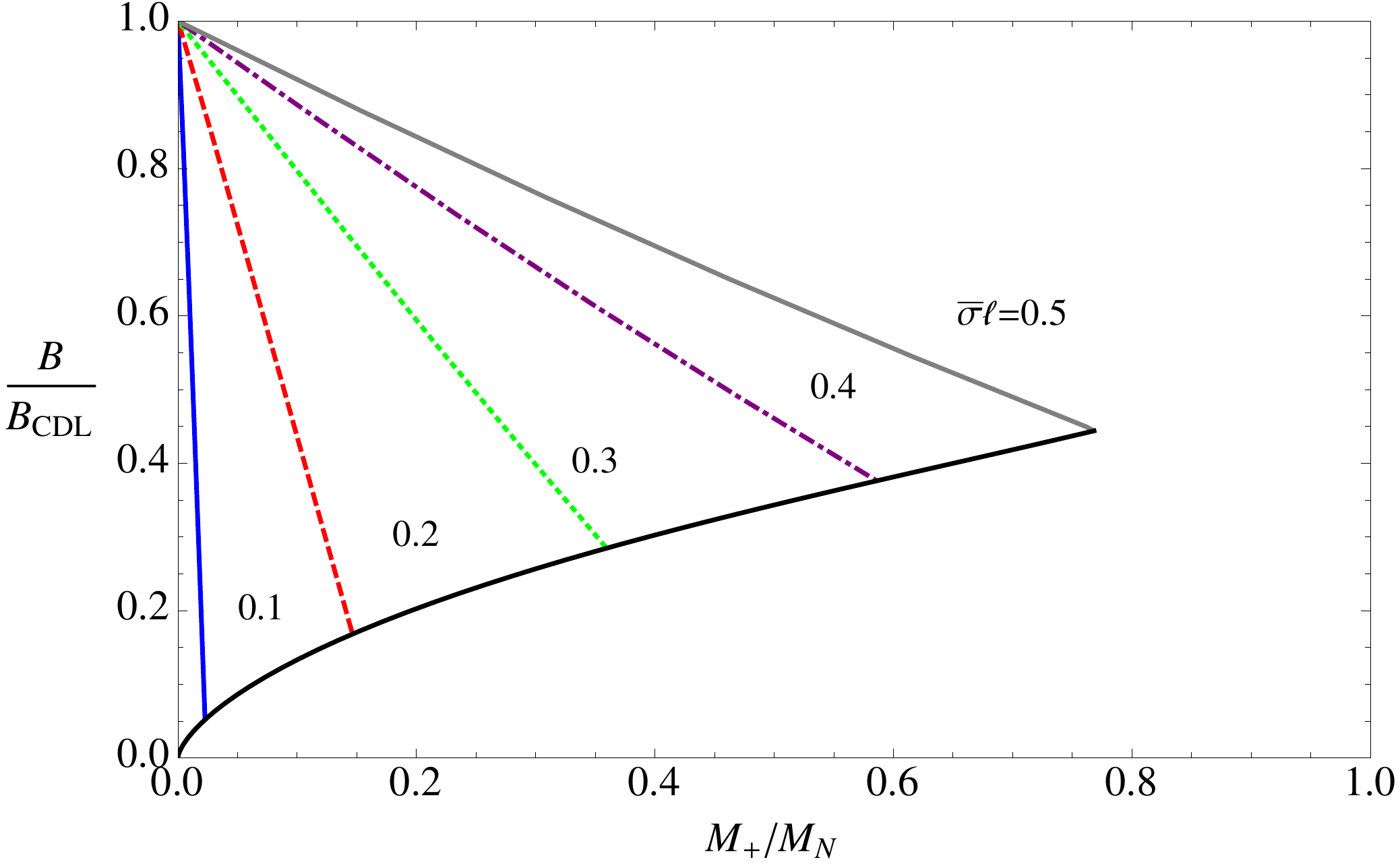}
\caption{The decay rate exponent $B$ for a Minkowski bubble, as given by 
\ceqref{decayrate}, as a function of the seed 
mass $M_+$ for fixed tension $\bar\sigma\ell$. The values are scaled by
$B_{CDL}$, the value of $B$ for the CDL instanton, and the Nariai
mass $M_N=\ell/\sqrt{27}G$. From left to right we have 
${\bar\sigma}\ell=0.1$ in solid blue, ${\bar\sigma}\ell=0.2$ 
in dashed red, ${\bar\sigma}\ell=0.3$ in dotted green, 
${\bar\sigma}\ell=0.4$ in dot-dash purple, and
${\bar\sigma}\ell=0.5$ in grey.  The lower solid black  curve 
indicates the `static' solutions, $\kappa_2=\kappa_\ast$,
which fixes a relationship between mass and tension.
In all cases, $B<B_{CDL}$ indicating that the decay rate is 
higher for a finite mass, at fixed tension.  }
\label{bfig}
}

\subsection{The critical static bubble wall with $M_-\ne 0$}
\label{insideblackhole}

The simplest situation in which a black hole remains 
behind after the tunnelling is the critical case where the masses
are determined by the value of $\bar\sigma\ell$ and the condition
$\kappa_1=\kappa_1^\ast(\kappa_2)$. For this critical bubble, we must solve 
the constraints \eqref{kappaconstraints}, then determine the decay rate
exponent for the `static' bounce, which is given by
\be
B_\ast = \frac{\pi(r_h^2 - (2GM_-)^2)}{G}.\label{bast}
\ee
Although all the expressions for $\kappa_1^\ast$, $r_h(GM_+)$
are algebraic, their form is not particularly illuminating (though we present
them in an appendix for completeness). We find numerically that the bounce 
action is always positive and minimal when $\kappa_2$ is maximal, for all 
values of ${\bar\sigma}\ell$. Naively, we might expect from \eqref{bast} that large 
remnant masses would have the smallest values of $B_*$. Paradoxically, we find 
that tardis like solutions, where the mass is bigger on the inside, generally
have larger action than the CDL case.

\subsection{The dominant processes}

In the previous subsections \ref{cdlsec}-\ref{insideblackhole}, we 
have detailed the behaviour of the Euclidean instanton actions 
in various special cases. The generic bubble solution depends on 
the masses $M_+$ and $M_-$ as well as the tension parameter 
$\bar\sigma\ell$, and the action has to be evaluated numerically. 
The important question is which one of these bubble solutions represents 
the dominant physical process. Specifically, for a  given seed black hole 
mass, $M_+$, we wish to minimise $B/B_{CDL}$ with respect to the 
remnant mass, $M_-$.

We have not been able to prove analytically which bubble solutions give the 
dominant rate, though we are able to perform a comprehensive 
numerical investigation. 
Here, we present results at a single fixed tension, $\bar{\sigma}\ell = 0.2$, 
though the other cases are qualitatively the same. 
For these bubbles, the quantity $B$ is shown in figure \ref{fig:argument} 
for a selection of remnant masses $M_-$ over the full range of $M_+$. 
Clearly, where they exist, the static bubbles provide the dominant 
contribution at fixed $M_+$, and where they do not, the $M_-=0$ 
solutions dominate. These solutions are indicated by the dashed lines 
in the figure. 
These are the solutions which maximise the mass difference, $\kappa_2$.

\FIGURE{
\includegraphics[width=0.7\textwidth]{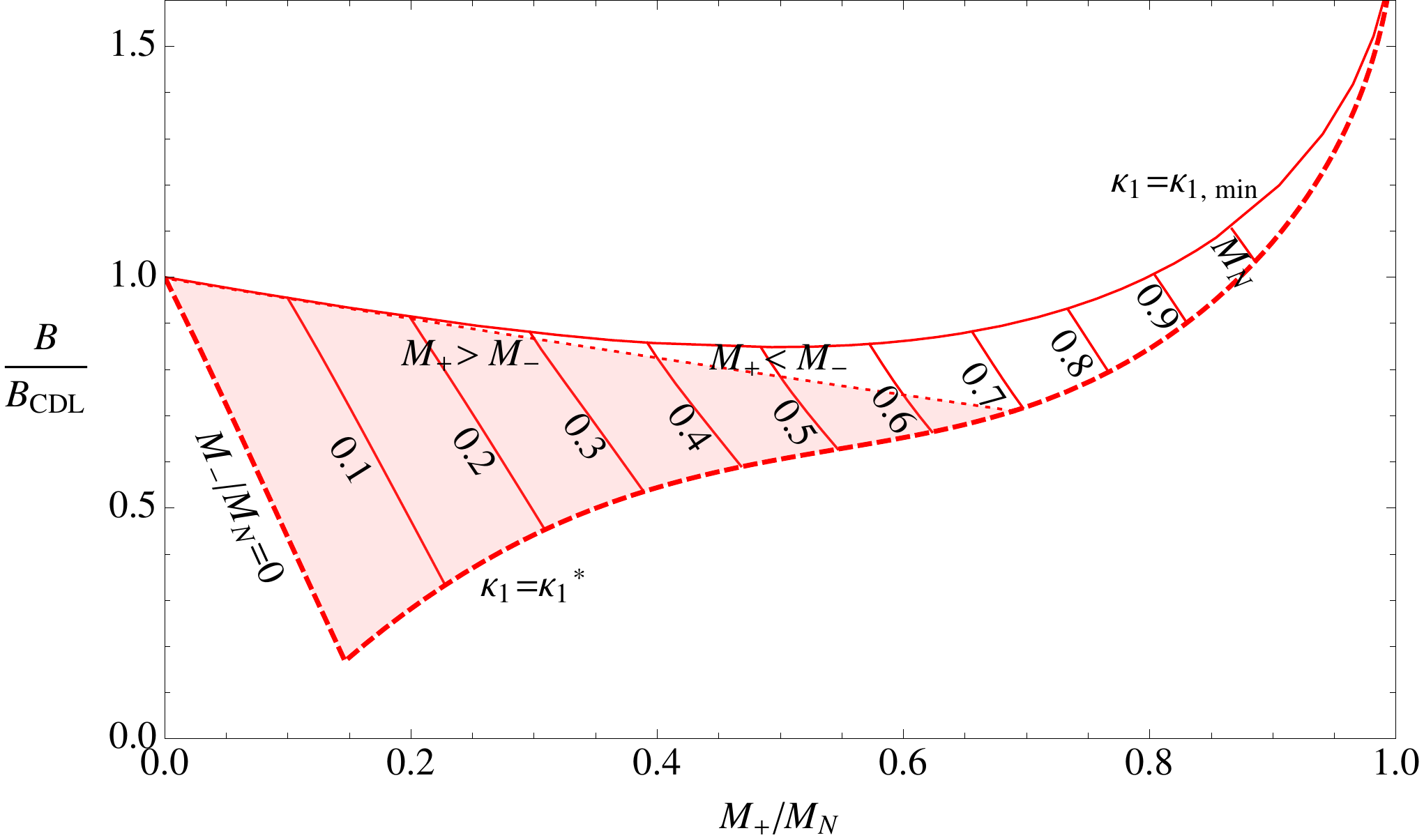}
\caption{A scan of the full parameter space of allowed Euclidean 
instantons at fixed tension $\bar\sigma\ell=0.2$. 
The upper red solid curve, together with the dashed lines give the full 
envelope of allowed solutions at this tension. 
Also indicated are lines of constant remnant mass, $M_-$, showing 
that the dominant solutions are given by $M_-=0$ where they exist i.e.  
$M_+<M_C$ with $M_C$ defined in \eqref{MCbulk}.
For $M_+>M_C$ this data shows that the dominant solutions 
are the static walls indicated by $\kappa_1=\kappa_1^\ast$.
The red shaded region corresponds to solutions where $M_-<M_+$, showing that for 
large enough $M_+$ the dominant processes become tardis-like.
}
\label{fig:argument}
}

Figure \ref{fig:argument} also indicates that the seed black hole which would 
give the highest rate process is given by a critical value, $M_C$ where, 
\be
G M_C = \frac{8}{27} \bar\sigma \gamma^2. \label{MCbulk}
\ee
as quoted earlier in \eqref{MCintro}. Extending the analysis to other 
tensions, we plot the value of $M_-$ which maximises the rate at a 
given $M_+$ in figure \ref{fig:dominantM}, and the corresponding 
values of $B$ in figure \ref{fig:dominantB}. Consistent bubble solutions
only exist when $\bar\sigma\ell<\sqrt{3}/2$ (see figure \ref{fig:kappas}).

\FIGURE{
\includegraphics[width=0.7\textwidth]{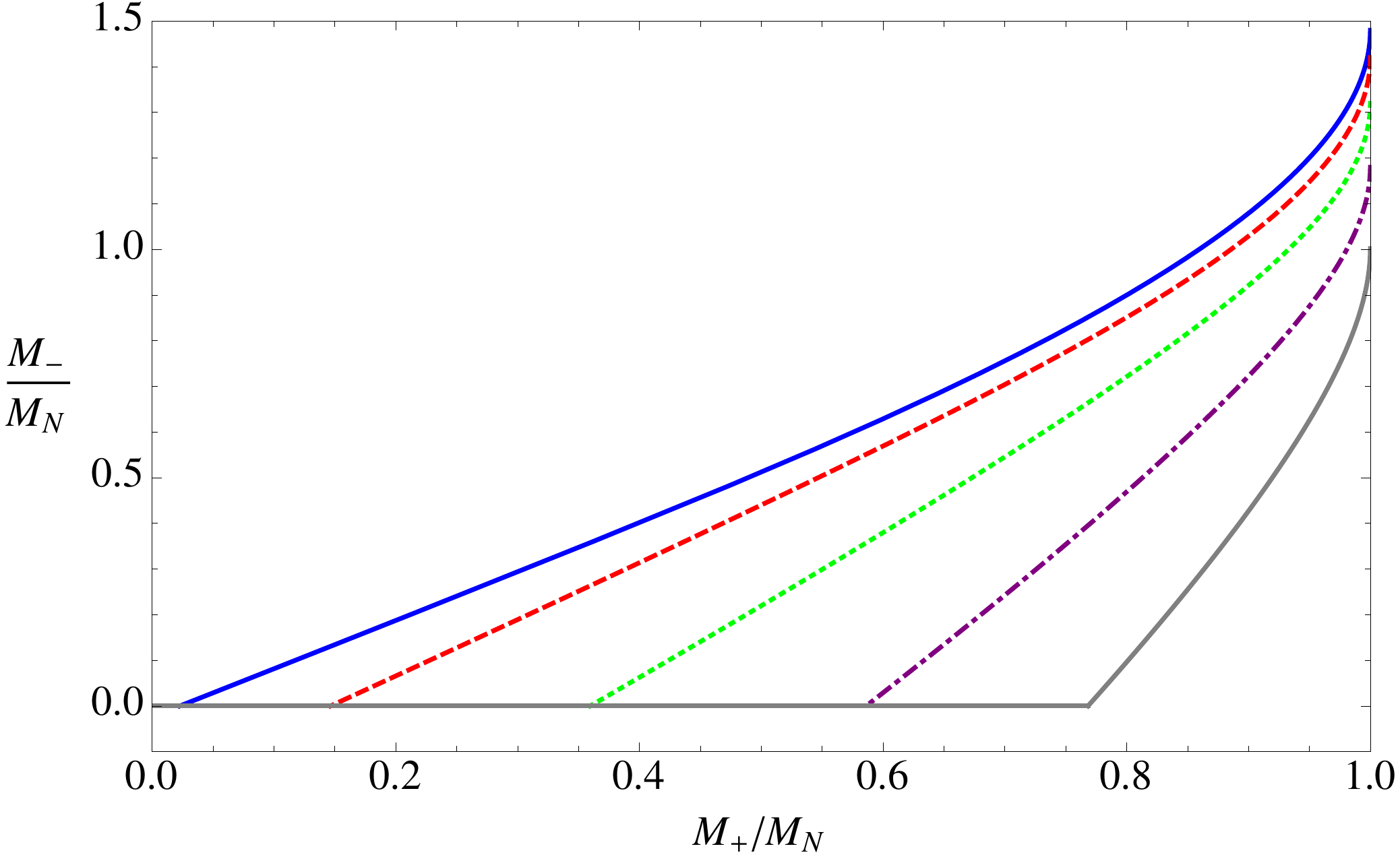}
\caption{The remnant mass $M_-$ as a function of the initial seed mass, $M_+$
with fixed values of the tension $\bar\sigma\ell$.
The remnant mass is only non-zero for sufficiently large seeds, $M_+>M_C$ .  
The results for ${\bar\sigma}\ell=0.1$ 
are shown in solid blue, ${\bar\sigma}\ell=0.2$ 
in dashed red, ${\bar\sigma}\ell=0.3$ in dotted green, 
${\bar\sigma}\ell=0.4$ in dot-dash purple, and
${\bar\sigma}\ell=0.5$ in gray. }
\label{fig:dominantM}
}

\FIGURE{
\includegraphics[width=0.7\textwidth]{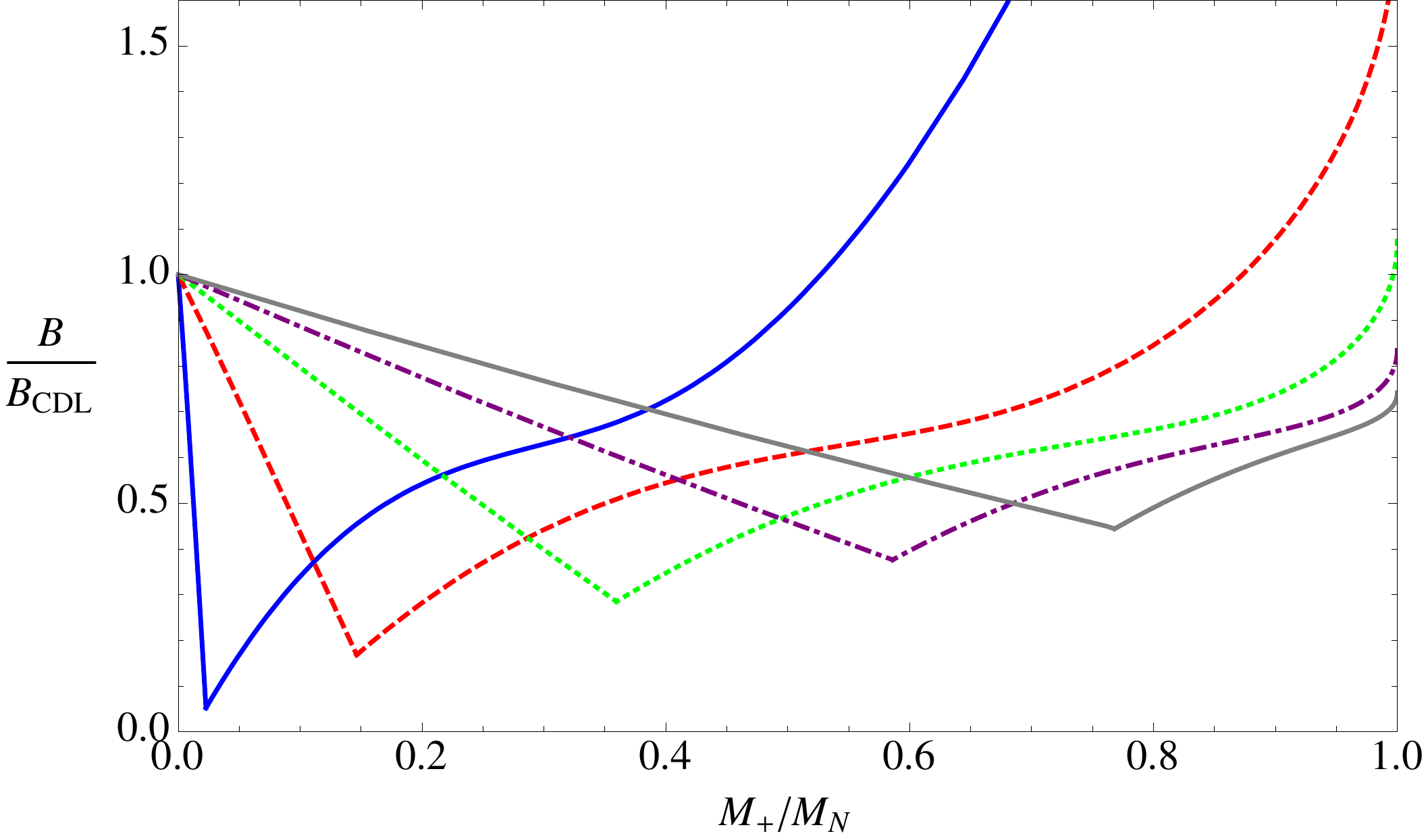}
\caption{The exponent in the decay rate for the dominant decay 
process as a function of the initial seed mass $M_+$ with fixed values 
of the tension $\bar\sigma\ell$. The value of $B$ is scaled by the CDL 
value and the mass by the Nariai mass $M_N=\ell/\sqrt{27}G$. 
The decay rate reaches its maximum value for the case of critical 
seed black holes $M_+=M_C$, as given by \eqref{MCbulk}. 
The results for ${\bar\sigma}\ell=0.1$ 
are shown in solid blue, ${\bar\sigma}\ell=0.2$ 
in dashed red, ${\bar\sigma}\ell=0.3$ in dotted green, 
${\bar\sigma}\ell=0.4$ in dot-dash purple, and
${\bar\sigma}\ell=0.5$ in gray. }
\label{fig:dominantB}
}

The underlying reason for the existence of this critical value is that the static 
walls only exist when $M_+\geq M_C$. Correspondingly, when $M_+> M_C$ 
the dominant process is the nucleation of a static bubble wall solution as 
discussed in section \ref{insideblackhole}. In this range the nucleation involves 
the creation of a black hole of finite mass $M_-$, indicated in figure 
\ref{fig:dominantM}. 
For $M_+> M_C$ the rate is therefore given by $B_\ast$ in \eqref{bast}.

When $M_+<M_C$, there is no longer a static wall solution and the 
maximum $\kappa_2$ solution has no black hole remnant, $M_-=0$. 
These solutions were previously explored in section \ref{generalm0bubbles} 
and $B$-values shown in figure \ref{bfig}.

\section{The WKB approach\label{wkb}}

The WKB approach was introduced by Fischler et al., 
\cite{Fischler:1989se,Fischler:1990pk}, to calculate the probability 
for transitions between various thin-walled bubble solutions. 
In this section we shall compare the WKB approach to the 
singular instanton approach we have been describing. 
The basic idea of the WKB method is to formulate an action 
which depends only on the bubble wall trajectory and then use 
the associated Schr\"odinger equation to calculate tunnelling 
probabilities. Fischler et al.\ began with the Einstein-matter action 
and used a coordinate system adapted to the bubble wall.
We introduce a far simpler version which uses the original 
static patch coordinates.

The gravity-matter action can be constructed in the same way 
as the Euclidean action discussion in Section \ref{actions}. The 
bubble wall is given by specifying an arbitrary function $R(\lambda)$. 
The background metric outside of the bubble wall is given by
the same metric used earlier (\ref{thesamemetric}), and as 
usual with spherically symmetric metrics, the time-independence of the 
functions $f_\pm(r)$ is a consequence of the constraint equations 
and independent of the wall trajectory. See \cite{BCG} for a fuller
discussion of this generalisation of Birkhoff's theorem, although note
the caveats of \cite{Charmousis:2003wm,Charmousis:2001nq} 
when additional matter is present. 
Repeating the steps (\ref{thesamemetric})-(\ref{generalaction}) 
in Lorentzian signature gives
\be
S=\frac{1}{16 \pi G}\int_{W}\left[ f^{-1}f_{,r}\rho\right]^+_-
+\frac{1}{8 \pi G}\int_{W}\left[K\right]^+_--\int_W\sigma
\ee
where $W$ is the bubble wall and
\be
\rho_\pm =(f_\pm +\dot R^2)^{1/2}.
\ee
The junction conditions determine the wall trajectory, and cannot be 
used at this stage, as we are seeking an effective action for this motion.
Instead, we work off-shell and evaluate the trace of extrinsic curvature 
using the earlier results,
\be
K=\rho^{-1}\ddot R+\frac12f_{,r}\rho^{-1}+2R^{-1}\rho.
\ee
After substituting the extrinsic curvature, the action can be written in 
terms of the Lagrangian of a one-dimensional dynamical system 
with coordinate $R(\lambda)$,
\be
S=\int L\,d\lambda
\ee
where
\be
L=\frac{1}{2G}\left[ \rho^{-1}\ddot R
+\frac12 f^{-1}f_{,r}\rho^{-1}\dot R^2+2\frac{\rho}{R}
\right]^+_- -4\pi\sigma R^2.
\ee
The classical equations of motion derived from this Lagrangian are 
second order but there is also a first order constraint which corresponds 
to relabelling of the coordinate $\lambda$ along the bubble wall. 
If we set $d\lambda=Nd\lambda'$ and then vary the action with respect 
to $N$, we find that the constraint reduces to the familiar junction 
condition\footnote{The orders of the field equation and the constraint 
are not obvious from an inspection of the Lagrangian.}
\be
\left[\rho\right]^+_-=4\pi G\sigma R.
\ee
As before, we may rewrite the constraint as a conservation law,
\be
\frac12\dot R^2+U=0
\ee
In the quantum theory, this constraint becomes an operator ${\cal H}$ 
and the operator constraint ${\cal H}\Psi=0$ acting on the wave 
function $\Psi$ becomes the  Schr\"odinger equation. Solutions can 
now tunnel through the barrier in the potential $V$. If we denote the 
tunnelling rate  by $\Gamma_{b\to b}$, then the WKB approximation gives,
\be
\Gamma_{b\to b}= e^{iS[R_b]}
\ee
where in this expression and in what follows the pre-factor to the 
exponential has been discarded. $R_b$ is the solution to the classical 
constraint with the complex `time' parameter $\lambda\to i\lambda$. 
If we compare  this to the instanton action $I$ evaluated earlier, 
by re-substituting $[K]$ back into the action, we notice that the 
contributions from the fixed points $r_h$ and $r_c$ are absent 
but otherwise $S$ is identical to $iI$. Consequently,
\be
\Gamma_{b\to b} = e^{-I-({\cal A}_-/4G)-({\cal A}_c/4G)},
\ee
where ${\cal A}_-$ is the area of the remnant black hole inside the bubble.
We stress again that this represents the tunnelling rate from bubble 
solutions to other bubble solutions. However, we can compare this to
the rate of false vaccum decay calculated using the instanton method,
\be
\Gamma = e^{-(I-I_{SdS})}=e^{-I-({\cal A}_+/4G)-({\cal A}_c/4G)}, 
\ee
where ${\cal A}_+$ is the area of the seed black hole nucleating the bubble.

\FIGURE{
\includegraphics[width=0.6\textwidth]{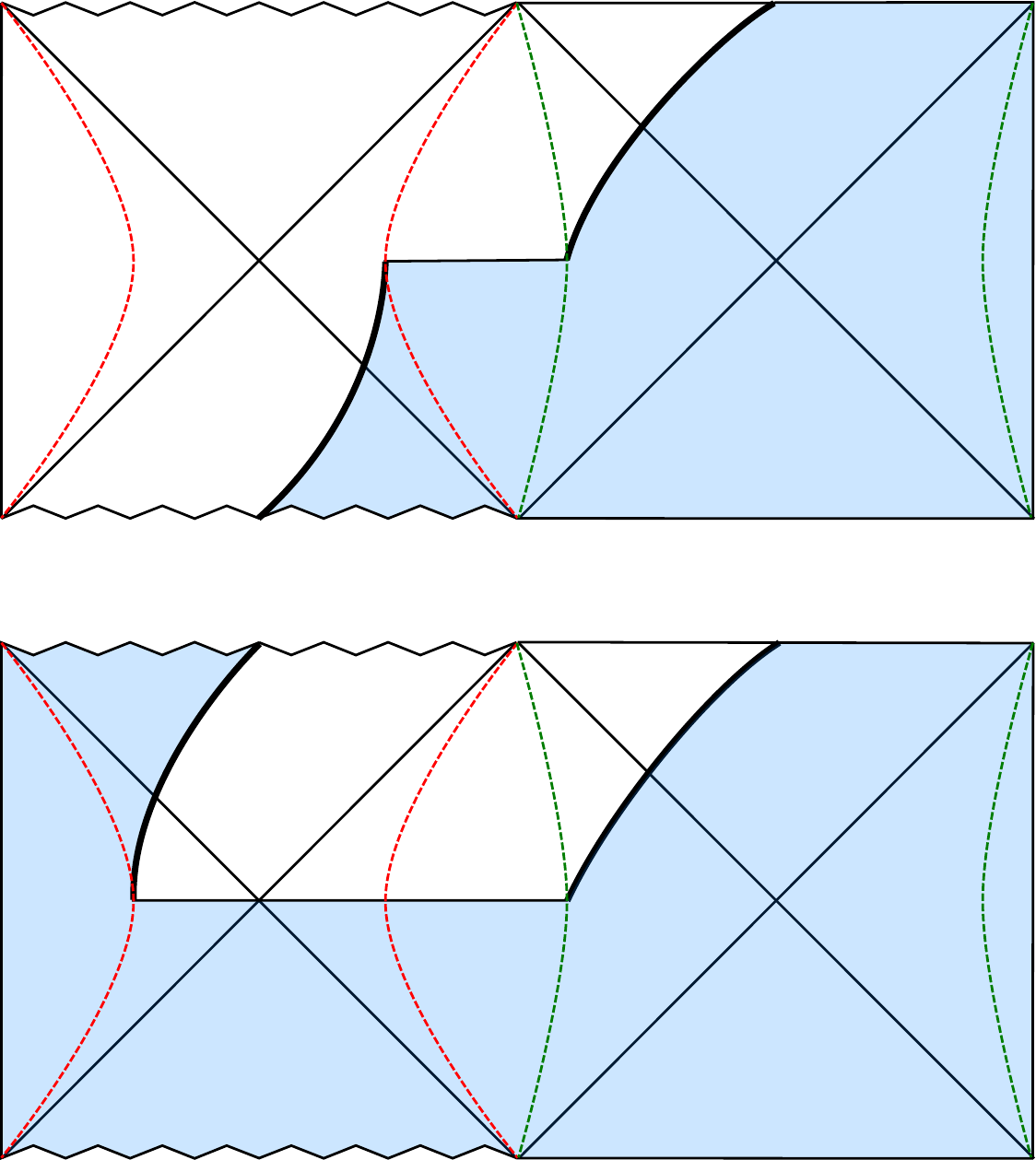}
\caption{These Penrose diagrams showing pieces of SdS illustrate 
the the similarities between bubble-bubble transitions and
bubble nucleations. The shaded regions represent the 
false vacuum. The bubble interiors are in the true vacuum with 
vanishing cosmological constant, and are not shown in the figures. 
In the first figure, the bubble wall starts out at $r = 0$, grows 
to $r = r_1$, tunnels to $r_2$ and grows again to $r=\infty$. 
This is interpreted as a bubble $\to$ bubble transition. 
In the second figure, two bubble walls are spawned from the 
vacuum, corresponding to vacuum $\to$ bubble $+$ bubble.
\label{crossingfig}}
}

On the fully extended SdS spacetime, the bubble wall is replicated in different
regions of the Penrose diagram, and so we might think of this as
the rate for the spontaneous production of two bubbles from the false vacuum 
$\Gamma_{f\to bb}$, as illustrated in figure \ref{crossingfig}.
Consequently, we have a relation of the form
\be
\Gamma_{f\to bb} = e^{-({\cal A}_+-{\cal A}_-)/4G} \,\Gamma_{b\to b}. 
\ee
We interpret this as a type of crossing relation between spontaeous production
of bubbles and bubble transitions, with the relative factor depending only on the
black hole entropies.

\section{Discussion\label{discussion}}

In this paper we have explored the gravitational effect of 
inhomogeneities on false vacuum decay. Our main result is 
that the presence of the inhomogeneity generically acts to 
enhance the rate of true-vacuum bubble nucleation. The exception occurs 
only for seed black holes which approach the Nariai limit.
The enhanced nucleation rates in the presence of a black hole could 
rescue some particle models which might otherwise get stuck in an 
early metastable state and never be able to decay into a radiation 
dominated universe.

We constructed a one-parameter family of Euclidean instantons
at each fixed seed mass $M_+$. In general, the nucleation 
processes these instantons describe occur at a rate far higher than 
that of the $M_+=0$ or CDL process. In particular, for
the limit of small seed masses $M_+\ll M_N$ and 
small tensions $\bar\sigma\ell\ll1$, the decay
exponent $B/B_{CDL}$ can be brought arbitrarily close to zero. 
The dominant decay at a fixed tension $\bar\sigma\ell$ depends on 
whether or not $M_+$ exceeds a critical value, $M_C$. 
For $M_+>M_C$ the decay 
also nucleates a remnant black hole inside the vacuum bubble. 
For $M_+<M_C$ the decay simply nucleates the vacuum bubble with a
flat interior.

The salient technical point of our analysis is the consistent 
treatment of conical singularities in the Euclidean instanton 
calculation\footnote{These singularities were not taken into
account in previous work \cite{PhysRevD.35.1161}.}, 
which faithfully reproduce existing results which have been obtained using 
manifestly regular constructions. For instance, in section \ref{cdlsec} 
we presented a derivation of the CDL instanton action starting 
from the Euclidean continuation of a single causal patch of de Sitter, 
containing conical singularities. 

For the massless case $M_-=M_+= 0$, and the critical mass case 
$\kappa_2 = \kappa_2^\ast$ with $M_-=0$, equivalent regular 
instanton constructions are known. In these two cases it seems 
that the treatment of conical 
singularities we have employed can be considered a proxy for the 
existence of a regular construction in a different Euclidean section. 
It would be interesting to investigate whether this is the case in 
general, i.e., whether there is a family of regular instantons for 
the general case of $\kappa_1,\kappa_2$.

We have also found that the bubble nucleation rate calculated by the
instanton approach is related to the bubble-to-bubble transition rate 
obtained in a Lorentzian WKB approach. 
One interpretation of this result is that the Euclidean instanton calculation 
actually describes the production of two true-vacuum bubbles in the fully extended
Schwarzshild-de Sitter spacetime, only one of which is present in a single static patch. 
The connection with the bubble-to-bubble transition rate can then be 
explained as a type of crossing relation between bubble-to-bubble 
and the production of two bubbles from the false vacuum.

In this analysis we employed the simplest nucleation seed -- 
a black hole. This may be introduced as an approximation to a 
spherical lump of matter, as shown by $S$ in figure \ref{collapsefig}.  
The space-time outside of the matter is made up of a finite number 
of the regions from the fully extended SdS space-time. 
There is no longer any need to include a black hole at the antipodal 
point of the universe, and only one bubble nucleates around the lump
of matter. Furthermore, the black hole approximation fails in a small 
region around the horizon, thus avoiding problems which would 
otherwise arise in defining stationary quantum states on the SdS 
background.

\FIGURE{
\includegraphics[width=0.6\textwidth]{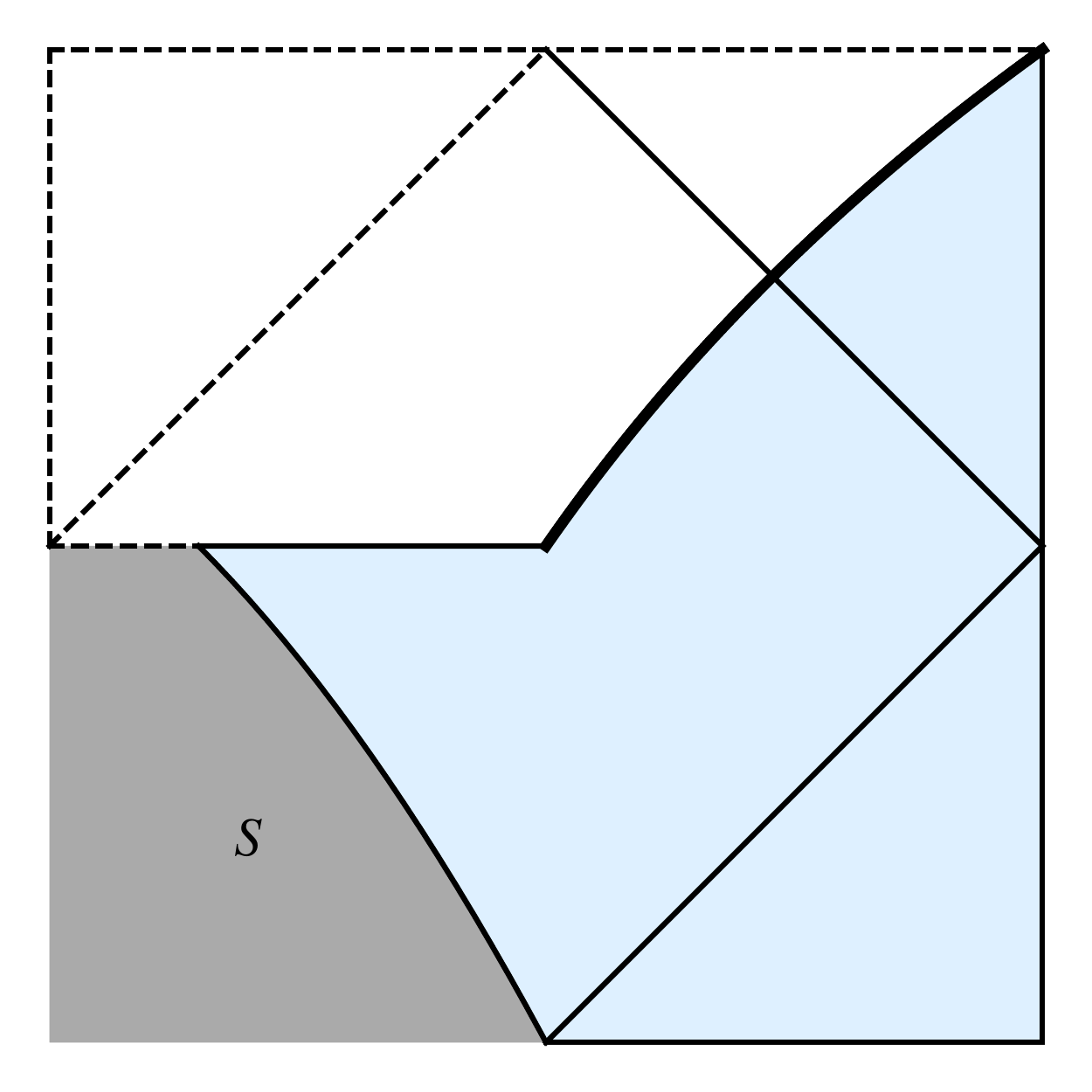}
\caption{The Penrose diagram for a collapsing lump of matter $S$ in 
the false vacuum region. The space-time outside of the collapsing lump 
is made up of just a finite number of regions from the fully extended 
space-time. }
\label{collapsefig}
}

An important omission in this paper has been the neglect of Hawking
radiation which may cause the black hole to evaporate before the vacuum
decay can occur. Some idea of the relative decay rates can be obtained
for black holes which are small compared to the horizon size. The largest
enhancement of the vacuum decay rate occurs for a black hole with both
$M_-=0$ and $M_+$ at the critical value, $M_C$, when the decay rate is
\be
\Gamma_\ast=A_\ast e^{-B_\ast},
\ee
where we have included the pre-factor $A_\ast$. 
From the results of section \ref{criticalresults}
we have $B_\ast=4\pi G M_C^2$. According to Callan 
and Coleman, \cite{callan1977}, this pre-factor is made up 
from a factor of $(B_\ast/2\pi)^{1/2}$ for each translational zero 
mode of the instanton and a determinant factor. 
In our case, there will be a single zero mode representing 
the time translation symmetry. Rather than evaluate the 
determinant factor, we use the inverse horizon timescale 
as a rough estimate $(GM_C)^{-1}$,  then
\be
\Gamma_\ast\approx \left(\frac{2}{G}\right)^{1/2}e^{-4\pi GM_C^2}.
\ee
The black hole emits Hawking radiation at a rate depending 
on fundamental particle masses and spins. The total decay rate 
for a subset of the standard model was evaluated by Page, 
\cite{Page}. If we set $\Gamma_H=\dot M/M$, then
\be
\Gamma_H\approx 3.6\times 10^{-4}(G^2 M_C^3)^{-1}
\ee
The ratio of the two decay rates is
\be
\frac{\Gamma_\ast}{\Gamma_H}\approx 3.9\times 10^3 
(GM_C^2)^{3/2}e^{-4\pi GM_C^2}.
\ee
Since this ratio is small for black holes larger than the Planck 
mass $GM_C^2>1$, these black holes decay before they 
can nucleate the false vacuum decay. The enhanced rate is 
relevant mostly for masses $M\ll M_C$ or for cases where 
the Hawking evaporation is suppressed, for example by de 
Sitter background radiation. Alternately, another physical mechanism,
such as accretion of a slowly rolling scalar, \cite{Chadburn:2013mta}, 
could negate the evaporative process (although this would 
simultaneously cause the tunnelling rate to drop).

Another interesting generalization of our study would be to explore
the effect in extra dimensional scenarios, such as
\cite{ArkaniHamed:1998rs,Randall:1999ee}, where the Planck
mass can drop substantially, leading to interesting new black
hole phenomenology, (for a review see \cite{Kanti:2004nr}). 
The CDL instanton has been generalised to a braneworld 
construction, \cite{Gregory:2001dn},
however, to include black holes in this picture would lead to the usual
impasse of a lack of an analytic exact solution for the C-metric
(see \cite{Gregory:2008rf} for a review of these issues, and
\cite{Figueras:2011gd} for recent numerical results).

Alternatively, there are many string theoretic models with `large'
or warped extra dimensions, such as KKLT, \cite{KKLT}, or LVS
(large volume scenarios, \cite{LVS}), in which a CDL type of
computation has been used to argue the long lifetime of a metastable
dS vacuum (see also \cite{AW,AGHQ}). For KKLT, the tension of the 
bubble wall lies outside the range allowed by the static patch 
construction: $2{\bar\sigma}\ell \gg 1$, \cite{KKLT}.
Thus we cannot directly apply our results, however, the 
intriguing possibility remains that the existence of a black hole 
could act as a nucleation site for decompactification. We leave the 
construction of supercritical black hole instantons for future
investigation.

\section*{Acknowledgements}
We would like to thank Patrick Dorey for discussions. 
RG and IM are supported in part by STFC (Consolidated Grant ST/J000426/1).
RG is also supported by the Wolfson Foundation and Royal Society, 
and by Perimeter Institute for Theoretical Physics. 
Research at Perimeter Institute is supported by the Government of
Canada through Industry Canada and by the Province of Ontario through the
Ministry of Research and Innovation.
BW was supported by a Royal Commission 
for the Exhibition of 1851 Research Fellowship.

\appendix
\section{Conical deficit regularisation}
\label{cdeficit}

In this appendix we review the computation of a conical
deficit action. Our assumption is that the metric has a specific product 
structure in which the conical deficit is in a 2-plane,
parametrized by local cylindrical coordinates, $\{\rho,\chi\}$,
and the transverse space $H$ is independent of
these coordinates as $\rho\to0$. Since we are interested in
near horizon geometries we will specify the metric to be
\be
ds^2 = d\rho^2 + A^2(\rho) d\chi^2 + 
C^2(\rho) d\Omega_H^2,
\ee
although the argument is independent of the precise structure
the sections transverse to $\{\rho,\chi\}$, provided $C'(0)=0$.
We define the area ${\cal A}$ of the conical defect  to be $C(0)^n$
times the area given by the metric $d\Omega_H^2$, where $n$
is the dimension of $H$.

The idea is now to smooth out the conical deficit by taking a
regular function $A$ such that $A'(0)=1$, $A'(\epsilon)=(1-\delta)$,
where $2\pi\delta$ is the deficit angle. Because $C(\rho)$
remains smooth, we may write $C=C_0 + \rho^2 C_2$, and hence
compute the Ricci scalar in the vicinity of $\rho=0$ as:
\be
\begin{aligned}
{\cal R} &= -\frac{2A''}{A} - \frac{2nC''}{C} - \frac{2nA'C'}{AC}+ 
\frac{n(n-1)(1-C'^2)}{C^2}\\
&\sim -\frac{2A''}{A} - \frac{4nC_2}{C_0} + \frac{n(n-1)}{C^2_0}
+ {\cal O} (\rho)
\end{aligned}
\ee
We see that this is the sum of a regular part (the terms
involving $C_0,C_2$ and ${\cal O}(\rho)$) and the
$A''/A$ term which becomes unbounded as $\epsilon\to0$,
as $A''= {\cal O}\left((A'(\epsilon)-A'(0))/\epsilon\right)$.
In computing the integral of the Ricci scalar over a small region
around $\rho=0$ therefore, it is only this unbounded term which
will contribute:
\be
\int d^n x \sqrt{g} \,{\cal R}
\sim {\cal A} [A'(0)-A'(\epsilon)]+ {\cal O} (\epsilon)
= 4\pi\delta\,{\cal A} + {\cal O} (\epsilon)
\ee
in agreement with \cite{Fursaev:1995ef}, for example.

To compute the Gibbons-Hawking boundary term, note that
the relevant inward pointing normal is $n = - d\rho$, with
extrinsic curvature $K=\nabla_a n^a = -A'/A - 2n C'/C$, hence
\be
\int_{\rho=\epsilon} d\chi d\Omega_H  AC^n K   \sim 
- 2 \pi {\cal A} \,A'(\epsilon) 
+ {\cal O} (\epsilon) = -2 \pi {\cal A} (1-\delta) + {\cal O} (\epsilon).
\ee

Combining these terms together, we see that the contribution of the
deficit angle, $\delta$, cancels, and taking $\epsilon\to0$,
we are left with the overall action:
\be
I_{\cal B} = 
-\int d^4x \sqrt{g} \frac{{\cal R}}{16\pi G} 
+ \int d^3 x \sqrt{h} \frac{K}{8\pi G}= -\frac{\cal A}{4G}
\ee

\section{The limits on $\kappa$}
\label{kappalimits}

The static solution is obtained when $U=U'=0$ for some ${\tilde R}_\ast$,
(${\tilde R}=R/\gamma$) which gives two polynomial constraints:
\be
\begin{aligned}
{\tilde R}_\ast^6 - \left ( \kappa_2+ \frac{\kappa_1}{2} \right ) 
{\tilde R}_\ast^3 - 2 \kappa_2^2 &=0 \\
{\tilde R}_\ast^3 - \frac23 {\tilde R}_\ast +  \kappa_2+ \frac{\kappa_1}{2}&=0
\end{aligned}
\label{kappaconstraints}
\ee
Consistency of the solution ${\tilde R}_\ast^3$ to the `quadratic', and 
${\tilde R}_\ast$ to the cubic then requires
\be
\begin{aligned}
\kappa_1 = \kappa_1^\ast &= \frac{1}{81} \Biggl [ 1 - \left ( 
-1 -5 (27\kappa_2)^2 + \frac{(27\kappa_2)^4}{2} + \frac{27\kappa_2}{2} 
\left ( 4 + (27\kappa_2)^2\right)^{3/2} \right)^{1/3}\\
& \qquad \;\;\;\;\; + \left ( 1 +5 (27\kappa_2)^2 - \frac{(27\kappa_2)^4}{2} 
+ \frac{27\kappa_2}{2} \left ( 4 + (27\kappa_2)^2\right)^{3/2} \right)^{1/3}
\Biggr]
\end{aligned}
\label{kappa1star}
\ee
which gives us an upper bound on $\kappa_1$.

To get a lower bound on $\kappa_1$, we use 
\be
f_+ {\dot\tau}_+ = \frac{\kappa_2}{{\tilde R}^2} + {\tilde R}
\left ( 1-2{\bar\sigma}\gamma\right) \geq 0
\ee
(the constraint from positivity of ${\dot \tau}_-$ being weaker).
This constraint is saturated when
${\tilde R}_+^3 = -\kappa_2/(1-2{\bar\sigma}\gamma)$. For 
${\bar\sigma}\ell>1/2$, we must have $\kappa_2>0$, and 
${\tilde R}_+$ must be greater than the maximum allowed
value of $\tilde R$, i.e.\ $U({\tilde R}_+), U'({\tilde R}_+) <0$.
Conversely, for ${\bar\sigma}\ell<1/2$, the nontrivial minimal
value for $\kappa_1$ occurs for $\kappa_2<0$, and ${\tilde R}_+$
must be less than the minimum allowed value of ${\tilde R}$,
i.e.\ $U({\tilde R}_+)<0$, $U'({\tilde R}_+)>0$. In each case, the range
closes off when ${\tilde R}_+ = {\tilde R}_\ast$, leading to the 
$\kappa-$limits:
\be
\begin{aligned}
\frac{4}{27} \geq \kappa_2 \geq \kappa_{2,min}(\bar\sigma) 
&= \frac{(2{\bar\sigma}\gamma-1)\ell^3}{3\sqrt{3}\gamma^3} \\
\kappa_1^\ast\geq
\kappa_1 \geq \kappa_{1,min} (\kappa_2,{\bar\sigma}) &= 
{\rm Max} \left \{ 
\frac{4\kappa_2 {\bar\sigma}^2\gamma^2}{(1-2{\bar\sigma}\gamma)}
+ \left | \frac{-\kappa_2}{(1-2{\bar\sigma}\gamma)}\right |^{1/3}
, 0 \right\}
\hskip -2mm
\end{aligned}
\label{kappamin}
\ee
From the range of $\kappa_2$, we conclude that ${\bar\sigma}\ell 
\leq \sqrt{3}/2$.

At the critical point $\kappa_{2,min}$, the seed mass
$M_+=M_N$, and the remnant mass $M_- = (3-4{\bar\sigma}^2\ell^2)
M_N/2$, hence the static bounce action is
\be
B_\ast = \frac{\pi \left [R_N^2 - (2GM_-)^2\right]}{G}
= \frac{8}{27} \frac{\pi \ell^2}{G} {\bar\sigma}^2 \ell^2
\left ( 3 - 2 {\bar\sigma}^2 \ell^2 \right)
\ee
Thus, although the bounce action does tend to zero as ${\bar\sigma}
\to0$, it does so far more slowly than $B_{CDL}$, which is 
proportional to ${\bar\sigma}^4$.

\providecommand{\href}[2]{#2}\begingroup\raggedright
\endgroup


\begin{thebibliography}{10}


\bibitem{coleman1977}
S.~Coleman, {\it {Fate of the false vacuum: Semiclassical theory}},  {\em
  Phys.Rev.} {\bf D15} (1977) 2929--36.

\bibitem{callan1977}
C. G. Callan and S.~Coleman, {\it {Fate of the false vacuum II: First quantum corrections}},  {\em
  Phys.Rev.} {\bf D16} (1977) 1762--68.

\bibitem{PhysRevD.23.347}
A.~H. Guth, {\it Inflationary universe: A possible solution to the horizon and
  flatness problems},  {\em Phys. Rev. D} {\bf 23} (Jan, 1981) 347--356.

\bibitem{1982Natur.298}
M.~S. Turner and F.~Wilczek, {\it {Is our vacuum metastable}},  {\em Nature}
  {\bf 298} (1982) 633.

\bibitem{PhysRevD.35.1161}
W.~A. Hiscock, {\it Can black holes nucleate vacuum phase transitions?},  
{\em  Phys. Rev. D} {\bf 35} (Feb, 1987) 1161--1170.

\bibitem{Berezin:1987ea}
V.~Berezin, V.~Kuzmin, and I.~Tkachev, {\it {O(3) invariant tunneling in
  general relativity}},  {\em Phys.Lett.} {\bf B207} (1988) 397.

\bibitem{PhysRevD.32.1333}
I.~G. Moss, {\it Black-hole bubbles},  {\em Phys. Rev. D} {\bf 32} (Sep, 1985)
  1333--1344.

\bibitem{Cheung:2013sxa}
C.~Cheung and S.~Leichenauer, {\it {Limits on New Physics from Black Holes}},
\href{http://xxx.lanl.gov/abs/1309.0530}{{\tt arXiv:1309.0530}}.

\bibitem{PhysRevD.21.3305}
S.~Coleman and F.~De~Luccia, {\it Gravitational effects on and of vacuum
  decay},  {\em Phys. Rev. D} {\bf 21} (Jun, 1980) 3305--3315.

\bibitem{Nar}
H.~ Nariai,
{\it On some static solutions of Einstein's gravitational field 
equations in a spherically symmetric case},
{\em Sci.\ Rept.\ Tohoku Univ.}\ {\bf 34}, 160 (1950);
{\it On a new cosmological solution of Einstein's field equations of gravitation},
{\em Sci.\ Rept.\ Tohoku Univ.}\ {\bf 35}, 46 (1951).

\bibitem{Hawking:1998bn}
S.~Hawking and N.~Turok, {\it {Open inflation without false vacua}},  
{\em Phys.Lett.} {\bf B425} (1998) 25--32,
[\href{http://xxx.lanl.gov/abs/hep-th/9802030}{{\tt hep-th/9802030}}].

\bibitem{Turok:1998he}
N.~Turok and S.~Hawking, {\it {Open inflation, the four form and the
  cosmological constant}},  {\em Phys.Lett.} {\bf B432} (1998) 271--278,
  [\href{http://xxx.lanl.gov/abs/hep-th/9803156}{{\tt hep-th/9803156}}].

\bibitem{gh1979}
G.~W. Gibbons and S.~W. Hawking, {\it {Classification of gravitational
  instanton symmetries}},  {\em Comm. Math. Phys.} {\bf 66} (1979) 291--310.

\bibitem{Garriga:2004nm}
J.~Garriga and A.~Megevand, {\it {Decay of de Sitter vacua by thermal
  activation}},  {\em Int.J.Theor.Phys.} {\bf 43} (2004) 883--904,
  [\href{http://xxx.lanl.gov/abs/hep-th/0404097}{{\tt hep-th/0404097}}].

\bibitem{Farhi:1989yr}
E.~Farhi, A.~H. Guth, and J.~Guven, {\it {Is it possible to create a universe
  in the laboratory by quantum tunneling?}},  {\em Nucl.Phys.} {\bf B339}
  (1990) 417--490.

\bibitem{Fischler:1989se}
W.~Fischler, D.~Morgan, and J.~Polchinski, {\it {Quantum nucleation of false
  vacuum bubbles}},  {\em Phys.Rev.} {\bf D41} (1990) 2638.

\bibitem{Fischler:1990pk}
W.~Fischler, D.~Morgan, and J.~Polchinski, {\it {Quantization of false vacuum
  bubbles: a Hamiltonian treatment of gravitational tunneling}},  {\em
  Phys.Rev.} {\bf D42} (1990) 4042--4055.

\bibitem{Aguirre:2005xs} 
A.~Aguirre and M.~C.~Johnson,
{\it Dynamics and instability of false vacuum bubbles},
{\em Phys.\ Rev.\ D} {\bf 72}, 103525 (2005)
[\href{http://xxx.lanl.gov/abs/gr-qc/0508093}{{\tt gr-qc/0508093}}].

\bibitem{Aguirre:2005nt}
A.~Aguirre and M.~C. Johnson, {\it {Two tunnels to inflation}},  {\em
  Phys.Rev.} {\bf D73} (2006) 123529,
  [\href{http://xxx.lanl.gov/abs/gr-qc/0512034}{{\tt gr-qc/0512034}}].



\bibitem{Israel}
W.~Israel,
{\it Singular hypersurfaces and thin shells in general relativity},
{\em Nuovo Cimento Soc.\ Ital.\ Phys.\ B}  {\bf 44}, 4349 (1966).

\bibitem{BCG}
P.~Bowcock, C.~Charmousis and R.~Gregory,
{\it General brane cosmologies and their global spacetime structure},
{\em Class.\ Quant.\ Grav.\ } {\bf 17}, 4745 (2000)
 [\href{http://xxx.lanl.gov/abs/hep-th/0007177}{{\tt  hep-th/0007177}}].
  
\bibitem{Mellor:1989gi}
F.~Mellor and I.~Moss, {\it {Black Holes and Quantum Wormholes}},  
{\em  Phys.Lett.} {\bf B222} (1989) 361.

\bibitem{Mellor:1989wc}
F.~Mellor and I.~Moss, {\it {Black Holes and Gravitational Instantons}},  
{\em  Class.Quant.Grav.} {\bf 6} (1989) 1379.

\bibitem{Dowker:1993bt}
F.~Dowker, J.~P. Gauntlett, D.~A. Kastor, and J.~H. Traschen, {\it {Pair
  creation of dilaton black holes}},  {\em Phys.Rev.} {\bf D49} (1994)
  2909--2917, [\href{http://xxx.lanl.gov/abs/hep-th/9309075}{{\tt
  hep-th/9309075}}].

\bibitem{Brown:2007sd}
A.~R. Brown and E.~J. Weinberg, {\it {Thermal derivation of the Coleman-De
  Luccia tunneling prescription}},  {\em Phys.Rev.} {\bf D76} (2007) 064003,
  [\href{http://xxx.lanl.gov/abs/0706.1573}{{\tt arXiv:0706.1573}}].

\bibitem{GT}
R.~P.~Geroch and J.~H.~Traschen,
{\it Strings and Other Distributional Sources in General Relativity},
{\em Phys.\ Rev.\ D} {\bf 36}, 1017 (1987)

\bibitem{Hawking:1995fd}
S.~Hawking and G.~T. Horowitz, 
{\it {The Gravitational Hamiltonian, action,  entropy and surface terms}},  
{\em Class.Quant.Grav.} {\bf 13} (1996) 1487--1498, 
[\href{http://xxx.lanl.gov/abs/gr-qc/9501014}{{\tt gr-qc/9501014}}].

\bibitem{Fursaev:1995ef}
D.~V. Fursaev and S.~N. Solodukhin, 
{\it {On the description of the Riemannian geometry in the 
presence of conical defects}},  
{\em Phys.Rev.} {\bf D52} (1995) 2133--2143, 
[\href{http://xxx.lanl.gov/abs/hep-th/9501127}{{\tt hep-th/9501127}}].

\bibitem{Page}
D. N. Page, {\it {Particle emission rates from a black hole}},  
{\em  Phys.Rev.} {\bf D13} (1976) 198-206.

\bibitem{Charmousis:2003wm} 
C.~Charmousis and R.~Gregory,
{\it Axisymmetric metrics in arbitrary dimensions},
{\em Class.\ Quant.\ Grav.\ } {\bf 21}, 527 (2004) 527--554
[\href{http://xxx.lanl.gov/abs/gr-qc/0306069}{{\tt gr-qc/0306069}}].
  
\bibitem{Charmousis:2001nq} 
C.~Charmousis,
{\it Dilaton space-times with a Liouville potential},
{\em Class.\ Quant.\ Grav.\  }{\bf 19}, 83 (2002) 83--114
[\href{http://xxx.lanl.gov/abs/hep-th/0107126}{{\tt hep-th/0107126}}].

\bibitem{Chadburn:2013mta} 
S.~Chadburn and R.~Gregory,
{\it Time dependent black holes and scalar hair},
\href{http://xxx.lanl.gov/abs/1304.6287}{{\tt arXiv:1304.6287}}.

\bibitem{ArkaniHamed:1998rs} 
N.~Arkani-Hamed, S.~Dimopoulos and G.~R.~Dvali,
{\it The Hierarchy problem and new dimensions at a millimeter},
{\em  Phys.\ Lett.\ B }{\bf 429}, 263 (1998) 263--272
[\href{http://xxx.lanl.gov/abs/hep-ph/9803315}{{\tt hep-ph/9803315}}].
  
\bibitem{Randall:1999ee} 
L.~Randall and R.~Sundrum,
{\it A Large mass hierarchy from a small extra dimension},
{\em Phys.\ Rev.\ Lett.\  }{\bf 83}, 3370 (1999) 3370--3373
[\href{http://xxx.lanl.gov/abs/hep-ph/9905221}{{\tt hep-ph/9905221}}].

\bibitem{Kanti:2004nr} 
P.~Kanti,
{\it Black holes in theories with large extra dimensions: A Review},
{\em Int.\ J.\ Mod.\ Phys.\ A }{\bf 19}, 4899 (2004) 4899-4951
[\href{http://xxx.lanl.gov/abs/hep-ph/0402168}{{\tt hep-ph/0402168}}].
  
\bibitem{Gregory:2001dn} 
R.~Gregory and A.~Padilla,
{\it Brane world instantons},
{\em Class.\ Quant.\ Grav.\  }{\bf 19}, 279 (2002)
[\href{http://xxx.lanl.gov/abs/hep-th/0107108}{{\tt hep-th/0107108}}].

\bibitem{Gregory:2008rf} 
R.~Gregory,
{\it Braneworld black holes},
{\em Lect.\ Notes Phys.\  }{\bf 769}, 259 (2009)
\href{http://xxx.lanl.gov/abs/0804.2595}{{\tt arXiv:0804.2595}}.

\bibitem{Figueras:2011gd} 
P.~Figueras and T.~Wiseman,
{\it Gravity and large black holes in Randall-Sundrum II braneworlds},
{\em  Phys.\ Rev.\ Lett.\  }{\bf 107}, 081101 (2011)
\href{http://xxx.lanl.gov/abs/1105.2558}{{\tt arXiv:1105.2558}}.

\bibitem{KKLT}
S.~Kachru, R.~Kallosh, A.~D.~Linde and S.~P.~Trivedi,
{\it De Sitter vacua in string theory},
{\em Phys.\ Rev.\ D} {\bf 68}, 046005 (2003)
[\href{http://xxx.lanl.gov/abs/hep-th/0301240}{{\tt hep-th/0301240}}].

\bibitem{LVS}
V.~Balasubramanian, P.~Berglund, J.~P.~Conlon and F.~Quevedo,
{\it Systematics of moduli stabilisation in Calabi-Yau flux compactifications},
{\em JHEP} {\bf 0503}, 007 (2005)
[\href{http://xxx.lanl.gov/abs/hep-th/0502058}{{\tt hep-th/0502058}}].

\bibitem{AW}
A.~Westphal,
{\it Lifetime of Stringy de Sitter Vacua},
{\em JHEP} {\bf 0801}, 012 (2008)
\href{http://xxx.lanl.gov/abs/0705.1557}{{\tt arXiv:0705.1557}}.

\bibitem{AGHQ}
S.~de Alwis, R.~Gupta, E.~Hatefi and F.~Quevedo,
{\it Stability, Tunneling and Flux Changing de Sitter Transitions 
in the Large Volume String Scenario},
{\em JHEP} {\bf 1311}, 179 (2013)
\href{http://xxx.lanl.gov/abs/1308.1222}{{\tt arXiv:1308.1222}}.



\end{thebibliography}
\end{document}